\documentclass[onecolumn,
prd,aps,tightenlines]{revtex4}

\begin{document}

\title{Pentaquark Decay Amplitudes From $SU(3)$ Flavor Symmetry}

\author{Steven M. Golbeck and Mikhail A. Savrov}
\affiliation{Department of Physics, University of California at San Diego,
  La Jolla, CA 92093\vspace{4pt} }
\date{\today}

\vskip1.5in
\begin{abstract}
The experimental signature of an $S=+1$ baryon of mass 1540 MeV and width $\sim 10-25$ MeV has led to  speculation that this is an exotic baryon consisting of $4$ quarks and $1$ antiquark lying at the top of a $\overline{10}$ $SU(3)_{F}$ multiplet. The observed decay modes are $\Theta^{+} \to K^{+}n$ and $\Theta^{+} \to K^0_Sp$ or in terms of flavor multiplets, $\overline{10} \to 8 \otimes 8$. The effective Hamiltonian for this decay can be written as a flavor singlet piece plus an $SU(3)$ breaking term. By using the Wigner-Eckart thereom, we then relate the decay amplitudes of all 10 members of the antidecuplet in terms of $5$ parameters. Similarly, we perform the same calculation for the proposed exotic multiplets $27$ and $35$. 
\end{abstract}

\maketitle

\section{Motivation}
The observed decay modes of $\Theta^{+} \rightarrow K^{+}n$ \cite{LEPS03,DIANA03,CLAS03-b,SAPHIR03,CLAS03-c,CLAS03-d} and $\Theta^{+} \rightarrow K^0_Sp$ \cite{ADK03,HERMES-04,SVD04,AER04,Salur04} have $S=+1$ and baryon number $B=+1$. As such, an ordinary $3$ quark baryon cannot be identified with the initial state. Instead, we are led to consider the possibility of an exotic baryon consisting of $4$ quarks and $1$ antiquark, which in this case would be an $\overline{s}$. The observed state has zero isospin and positive charge, so it may be part of an $I=1$ isomultiplet. Searches for higher isopartners, however, have thus far been negative. All the data therefore points to this being an $I=0$ exotic baryon with $S=+1$, and $SU(3)_F$ gives the antidecuplet $\overline{10}$ as a probable candidate for its flavor multiplet. However the possibility that the state actually belongs to the $27$ or $35$ mutiplet  has not yet been ruled out. There has been much activity in explaining the observed resonance in chiral soliton models \cite{diakonov, Praszalowicz, Praszalowicz:2003tc, Itzhaki:2003nr, Cohen:2003mc, manohar} and the diquark model \cite{jaffe}. Work has also been done in exploiting the $SU(3)_F$ symmetry among the light quarks to determine pentaquark decay amplitudes \cite{schat,lee1, oh}. Here we adopt a similar approach to the latter, but include $SU(3)_F$ breaking in the analysis.

$SU(3)_F$ has proven to be a very accurate symmetry in the baryon sector of QCD bound states due to the small mass differences between the u,d, and s quark masses relative to $\Lambda_{QCD}$. $SU(3)_F$ breaking can be incorporated through the $T^{8}$ Gell-Mann matrix which treats the s quark differently than the u and d quarks, and if further accuracy is required, isospin breaking can be generated via the $T^{3}$ matrix. 

\section{The Effective Hamiltonian}\label{Hamiltonian}
If, in fact, the observed decay mode is that of an exotic baryon lying at the top of an antidecuplet, then we can parametrize all the decay modes of its members by using $SU(3)_F$ symmetry and incorporating $SU(3)_F$ breaking to first order in our effective Hamiltonian:
\begin{equation}
H = \alpha \openone +  {\cal{O}}^{i}_{j} [T^8]^{j}_{i}
\label{eq:heff}
\end{equation}
where we have explicitly shown the $SU(3)_F$ transformation properties of each term in the Hamiltonian. Since Isospin splittings are of the order of a few MeV, we have neglected an $SU(2)$ breaking term in our Hamiltonian corresponding to this effect.

We wish to calculate the matrix element:
\begin{equation}
\langle \mathbf{8} \otimes \mathbf{8} | H | \overline{\mathbf{10}} \rangle
\end{equation}
which will yield the amplitude for a pentaquark to decay into a pseudoscalar meson and octet baryon.

According to the Wigner-Eckart thereom, we can separate out the dynamical and group theoretical pieces of the matrix element into a reduced matrix element in which the dynamics is encoded, and a matrix element that depends solely upon the $SU(3)$ properties of the states and Hamiltonian. This leads to:
\begin{eqnarray}
\langle \mathbf{8} \otimes \mathbf{8} | H | \overline{\mathbf{10}} \rangle &=& 
\alpha \langle ^{i}_{j}| \langle ^{k}_{l}   | ^{rst} \rangle (B^*)^{j}_{i} (M^*)^{l}_{k} D_{rst} +  \langle ^{i}_{j}| \langle ^{k}_{l} |  {\cal{O}}^{m}_{n}  | ^{rst} \rangle (B^*)^{j}_{i} (M^*)^{l}_{k} [T^8]^{n}_{m} D_{rst}
\label{eq:decay}
\end{eqnarray}
where we have represented the baryon octet, pseudoscalar meson octet, and antidecuplet states respectively as:
\begin{eqnarray}
\langle 8 | &=& \langle ^{i}_{j} | (B^*)^{j}_{i} \nonumber\\
\langle 8 | &=& \langle ^{k}_{l} | (M^*)^{l}_{k} \nonumber\\
| \overline{10} \rangle &=& D_{rst} |^{rst} \rangle \nonumber\\
\end{eqnarray}
with the baryon octet tensor coefficients:
\begin{eqnarray}
&&
B^1_3 = P , \hspace{2cm}
B^2_3 = N, \hspace{1.6cm}
B^1_2 = \Sigma^+ , 
\nonumber \\ &&
B^1_1 = \frac{ \Sigma^0}{\sqrt{2}} + \frac{\Lambda^0}{\sqrt{6}} , \quad
B^2_2 = -\frac{ \Sigma^0}{\sqrt{2}} + \frac{\Lambda^0}{\sqrt{6}} , \quad
B^3_3 = -\frac{2 \Lambda^0}{\sqrt{6}} , \nonumber \\ &&
B^2_1=\Sigma^- , \hspace{1.8cm}
B^3_2 = \Xi^0, \hspace{1.6cm}
B^3_1 = \Xi^-.
\label{eq:baryon}
\end{eqnarray}
The meson tensor coefficients are:
\begin{eqnarray}
&&
M^1_3 = K^+ , \hspace{1.8cm}
M^2_3 = K^0, \hspace{1.6cm}
M^1_2 = \pi^+ , 
\nonumber \\ &&
M^1_1 = \frac{ \pi^0}{\sqrt{2}} + \frac{\eta^0}{\sqrt{6}} , \quad
M^2_2 = -\frac{ \pi^0}{\sqrt{2}} + \frac{\eta^0}{\sqrt{6}} , \quad
M^3_3 = -\frac{2 \eta^0}{\sqrt{6}} , \nonumber \\ &&
M^2_1=\pi^- , \hspace{1.8cm}
M^3_2 = \overline{K}^0, \hspace{1.6cm}
M^3_1 = K^-.
\label{eq:meson}
\end{eqnarray}
and the antidecuplet has tensor coefficients:
\begin{eqnarray}
&&
D_{333} = \Theta_{\overline{10}}^+ , \qquad
D_{233} = \frac{P_{\overline{10}}}{\sqrt{3}}, \qquad
D_{133} = \frac{N_{\overline{10}}}{\sqrt{3}}, \qquad
\nonumber \\ &&
D_{223} = \frac{\Sigma^{+}_{\overline{10}}}{\sqrt{3}}, \qquad
D_{123} = \frac{\Sigma^{0}_{\overline{10}}}{\sqrt{6}} , \qquad
D_{113} = \frac{\Sigma^{-}_{\overline{10}}}{\sqrt{3}}, \qquad
\nonumber \\ &&
D_{222} = \Xi^{+}_{\overline{10}}, \qquad
D_{122} = \frac{\Xi^{0}_{\overline{10}}}{\sqrt{3}}, \qquad
D_{112} = \frac{\Xi^{-}_{\overline{10}}}{\sqrt{3}}, \qquad
\nonumber \\ &&
D_{111} = \Xi^{--}_{\overline{10}}
\end{eqnarray}
The indices of the $\overline{10}$ are completely symmetric and thereby we omitted redundant elements from the tensor above.

\section{Tensor Decomposition}\label{Tensor}
In order to make usage of our expression for the decay amplitude we need to know what irreducible representations of $SU(3)$ the tensor products $8 \otimes 8$ and $\overline{10} \otimes 8$, where the $ 8$ in the second tensor product corresponds to the $SU(3)$ breaking term in the Hamiltonian, decompose into:
\begin{eqnarray}
8 \otimes 8 &=& 1 \oplus 8 \oplus 8 \oplus 10 \oplus \overline{10} \oplus 27 \nonumber\\
\overline{10} \otimes 8 &=& 8 \oplus \overline{10} \oplus 27 \oplus \overline{35} \nonumber\\
\label{eq:decomp}
\end{eqnarray}
Because there is a $\overline{10}$ in the $8 \otimes 8$ decomposition, the pentaquark $\overline{10}$ will be able to decay into an octet baryon $+$ pseudoscalar meson via the singlet piece of the Hamiltonian. In addition, both $8 \otimes 8$ and $\overline{10} \otimes 8$ contain an $8, \overline{10},$ and $27$, so there will also be contributions from the $SU(3)$ breaking term. As there are two octets in $8 \otimes 8$, the total number of reduced matrix elements will be five, which correspond to one from the singlet and four from the $SU(3)$ breaking term, so we may then calculate all decay amplitudes of the ten members of the antidecuplet in terms of five parameters.

The decay amplitude is a scalar quantity, so all the indices on the tensors must be fully contracted. Because of the orthogonality of different irreducible representations, what indices we contract can be determined in a fairly straightforward fashion \cite{Georgi}.  The explicit form for the  $8,\overline{10}$,and $27$ tensors from each tensor product is then necessary. The fundamental form of each of these for $8 \otimes 8$ are:
\begin{eqnarray}
(8_a)^{i}_{j} &=& B^{i}_{m}M^{m}_{j} - \frac{1}{3}\delta^{i}_{j} B^{n}_{m}M^{m}_{n} \nonumber\\
(8_b)^{i}_{j} &=& B^{m}_{j}M^{i}_{m} - \frac{1}{3}\delta^{i}_{j} B^{n}_{m}M^{m}_{n} \nonumber\\
\overline{10}_{ijk} &=& \epsilon_{irs}(B^{r}_{j}M^{s}_{k}+B^{r}_{k}M^{s}_{j}) + \epsilon_{jrs}(B^{r}_{k}M^{s}_{i}+B^{r}_{i}M^{s}_{k}) + \epsilon_{krs}(B^{r}_{i}M^{s}_{j}+B^{r}_{j}M^{s}_{i}) \nonumber\\
27^{ik}_{jl} &=& (B^{i}_{j}M^{k}_{l} + B^{i}_{l}M^{k}_{j} + B^{k}_{j}M^{i}_{l} + B^{k}_{l}M^{i}_{j}) - \frac{1}{5}[\delta^{i}_{l}(B^{m}_{j}M^{k}_{m} + B^{k}_{m}M^{m}_{j}) \nonumber\\
& +& \delta^{i}_{j}(B^{m}_{l}M^{k}_{m} + B^{k}_{m}M^{m}_{l}) + \delta^{k}_{l}(B^{m}_{j}M^{i}_{m} + B^{i}_{m}M^{m}_{j}) + \delta^{k}_{j}(B^{m}_{l}M^{i}_{m} + B^{i}_{m}M^{m}_{l})] \nonumber\\
& +& \frac{1}{10}[(\delta^{i}_{l} \delta^{k}_{j} + \delta^{i}_{j} \delta^{k}_{l})B^{m}_{n}M^{n}_{m}] \nonumber\\
\label{eq:8x8}
\end{eqnarray}
There are two different expressions for $8$ because it appears twice in the tensor decomposition of Eq. \ref{eq:decomp}. The same irreducible representations appearing in the tensor decomposition of $\overline{10} \otimes 8$ are:
\begin{eqnarray}
8^{i}_{j} &=& \epsilon^{irs}[T^{8}]^{n}_{r} D_{nsj} \nonumber\\
\overline{10}_{ijk} &=& [T^{8}]^{n}_{i} D_{njk} + [T^{8}]^{n}_{j} D_{nik} + [T^{8}]^{n}_{k} D_{nji} \nonumber\\
27^{ik}_{jl} &=& [\epsilon^{irs}[T^{8}]^{k}_{r} D_{sjl} + \epsilon^{krs}[T^{8}]^{i}_{r} D_{sjl}] - \frac{1}{5}[\delta^{i}_{l} \epsilon^{krs}[T^{8}]^{m}_{r} D_{sjm}
+ \delta^{i}_{j}\epsilon^{krs}[T^{8}]^{m}_{r} D_{slm} \nonumber\\
& & + \delta^{k}_{j}\epsilon^{irs}[T^{8}]^{m}_{r} D_{slm}
+ \delta^{k}_{l}\epsilon^{irs}[T^{8}]^{m}_{r} D_{sjm}] \nonumber\\
\label{eq:10bx8}
\end{eqnarray}
The $\overline{10}$ and $27$ have the required symmetry among their upper and lower indices resprectively, and the $8$ and $27$ tensors are also traceless as is necessary.

\section{The $\overline{10}$ Decay Amplitude}\label{Ten}
The decay amplitude may now be evaluated by contracting the irreducible tensors from $8 \otimes 8$ with the corresponding tensor from $\overline{10} \otimes 8$. There is a unique contraction for each irreducible representation. The tensor coefficients of a bra state vector are the complex-conjugate transpose of the tensor coefficients of the ket state vector, for example:
\begin{equation}
| \overline{10} \rangle = D_{rst} |^{rst} \rangle \to  \langle ^{rst}| (D^*)_{rst} = \langle \overline{10}|
\end{equation}
In terms of reduced matrix elements, the decay amplitude is:
\begin{eqnarray}
\langle \mathbf{8} \otimes \mathbf{8} | H | \overline{\mathbf{10}} \rangle &=& \alpha_{\overline{10}}* ({\overline{10}}^*)_{lmn}D_{lmn} + \beta_{\overline{10}}* (\overline{10}^*)_{lmn} {\overline{10}}_{lmn} + 
\gamma_{\overline{10}}* {(8_a^*)^{n}_{m}} {8^{n}_{m}}  \nonumber\\
& &+ \delta_{\overline{10}}* {(8_b^*)^{n}_{m}} {8^{n}_{m}} + \epsilon_{\overline{10}}* {(27^*)^{mn}_{kl}} {27^{mn}_{kl}}
\label{eq:decay1}
\end{eqnarray}
where the greek letters are the reduced matrix elements and the first irreducible representation in each term corresponds to the $8 \otimes 8$ tensors defined in Eq. \ref{eq:8x8}, whereas the second tensor corresponds to those of $\overline{10} \otimes 8$ defined in Eq. \ref{eq:10bx8}. Again their are two terms involving the octets as there are two octets in the $8 \otimes 8$ decomposition. To emphasize that this quantity is an $SU(3)$ scalar, we could have written it in terms of tensor coefficients $\overline{B}$ and $\overline{M}$ where:
\begin{equation}
\overline{B}^{i}_{j} = (B^*)^{j}_{i}
\end{equation}
which holds component wise. It is perhaps more transparent to use the RHS of the equality as we have explicit forms for these tensors in Eq. \ref{eq:8x8}.

The first term in Eq. \ref{eq:decay1} corresponds to the singlet piece in the Hamiltonian of Eq. \ref{eq:decay}, and is the dominant contribution to the amplitude. The four remaining terms are all from the $SU(3)$ breaking term in the Hamiltonian, and as such, one expects them to be on the order of one-third the strength of the singlet contribution. The coefficients of each reduced matrix element for all the decay modes of the $\overline{10}$ into an octet baryon + pseudoscalar meson are tabulated in Appendix \ref{table1}. The particle assignments are given in Appendix \ref{su3}.

\section{The $27$ Decay Amplitude}
\label{27}
Similarly to the calculation for the $\overline{10}$ decay amplitude outlined above, we may also consider the decays of the postulated $27$ exotic multiplet into octet baryons + pseudoscalar mesons. From Eq. \ref{eq:decomp}, we see that there is a $27$ in the decomposition of $8 \otimes 8$, so the $27$ exotic multiplet will decay into $8 \otimes 8$ via the singlet piece of the Hamiltonian. There will also be contributions from the $SU(3)$ breaking term in the Hamiltonian as we have:
\begin{eqnarray}
27 \otimes 8 &=& 8 \oplus 10 \oplus \overline{10} \oplus 27 \oplus 27 \oplus 35 \oplus \overline{35} \oplus 64
\label{eq:decomp1}
\end{eqnarray}
Referencing this against Eq. \ref{eq:decomp}, we see that there will be six reduced matrix elements for the $SU(3)$ breaking term, each corresponding to the matrix element between the same irreducible representations in the decompositions, and hence a total of seven parameters will govern the decay amplitude of the $27$. 

The $27$ state is:
\begin{equation}
|27\rangle = W^{ik}_{jl} |^{jl}_{ik}\rangle
\end{equation}
where $W^{ik}_{jl}$ is both traceless and symmetric in its upper and lower indices respectively with elements:

\medskip
\textbullet\ $S = 1, I=1$
\begin{eqnarray}
W^{11}_{33} = \Theta_{27}^{++}, \qquad
W^{12}_{33} = \frac{\Theta_{27}^{+}}{\sqrt{2}}, \qquad
W^{22}_{33} = \Theta_{27}^{0},
\end{eqnarray}

\textbullet\ $S = 0, I=3/2,1/2$
\begin{eqnarray}
&&
W^{11}_{23} = \frac{\Delta_{27}^{++}}{\sqrt{2}}, \qquad
W^{11}_{13} = \frac{\Delta_{27}^{+}}{\sqrt{6}} + \frac{2 P_{27}}{\sqrt{30}}, \qquad
W^{12}_{23} = -\frac{\Delta_{27}^{+}}{\sqrt{6}} + \frac{P_{27}}{\sqrt{30}}, \qquad
W^{13}_{33} = - \frac{3 P_{27}}{\sqrt{30}}, \qquad
\nonumber \\ &&
W^{12}_{13} = -\frac{\Delta_{27}^{0}}{\sqrt{6}} + \frac{N_{27}}{\sqrt{30}}, \qquad
W^{22}_{23} = \frac{\Delta_{27}^{0}}{\sqrt{6}} + \frac{2 N_{27}}{\sqrt{30}}, \qquad
W^{23}_{33} = -\frac{3 N_{27}}{\sqrt{30}}, \qquad
W^{22}_{13} = \frac{\Delta_{27}^{-}}{\sqrt{2}},
\end{eqnarray}

\textbullet\ $S = -1, I=2,1,0$
\begin{eqnarray}
&&
W^{11}_{22} = \Sigma_{27}^{'++}, \qquad
W^{11}_{12} = \frac{\Sigma_{27}^{'+}}{2} +  \frac{ \Sigma_{27}^+}{2\sqrt{5}}, \qquad
W^{12}_{22} = -\frac{\Sigma_{27}^{'+}}{2} +  \frac{ \Sigma_{27}^+}{2\sqrt{5}},  \qquad
W^{13}_{23} = - \frac{ \Sigma_{27}^+}{\sqrt{5}}, \qquad
\nonumber \\ &&
W^{11}_{11} = \frac{\Sigma_{27}^{'0}}{\sqrt{6}} - \frac{\Sigma_{27}^0}{\sqrt{10}} - \frac{\Lambda_{27}^0}{\sqrt{30}}, \qquad
W^{12}_{12} = -\frac{\Sigma_{27}^{'0}}{\sqrt{6}} - \frac{\Lambda_{27}^0}{2\sqrt{30}}, \qquad
W^{13}_{13} =  \frac{\Sigma_{27}^0}{\sqrt{10}} + \frac{3\Lambda_{27}^0}{2\sqrt{30}}, \qquad
\nonumber \\ &&
W^{22}_{22} = \frac{\Sigma_{27}^{'0}}{\sqrt{6}} + \frac{\Sigma_{27}^0}{\sqrt{10}} - \frac{\Lambda_{27}^0}{\sqrt{30}}, \qquad
W^{23}_{23} = - \frac{\Sigma_{27}^0}{\sqrt{10}} + \frac{3\Lambda_{27}^0}{2\sqrt{30}}, \qquad
W^{33}_{33} = - \frac{3\Lambda_{27}^0}{\sqrt{30}}, \qquad
\nonumber \\ &&
W^{12}_{11} = \frac{\Sigma_{27}^{'-}}{2} + \frac{\Sigma_{27}^-}{2\sqrt{5}}, \qquad
W^{22}_{12} = -\frac{\Sigma_{27}^{'-}}{2} + \frac{\Sigma_{27}^-}{2\sqrt{5}}, \qquad
W^{23}_{13} = -\frac{\Sigma_{27}^-}{\sqrt{5}}, \qquad
W^{22}_{11} = \Sigma_{27}^{'--},
\end{eqnarray}

\textbullet\ $S = -2, I=3/2,1/2$
\begin{eqnarray}
&&
W^{13}_{22} = \frac{\Xi_{27}^{'+}}{\sqrt{2}}, \qquad
W^{13}_{12} = \frac{\Xi_{27}^{'0}}{\sqrt{6}} + \frac{\Xi_{27}^{0}}{\sqrt{30}}, \qquad
W^{23}_{22} = -\frac{\Xi_{27}^{'0}}{\sqrt{6}} + \frac{2 \Xi_{27}^{0}}{\sqrt{30}}, \qquad
W^{33}_{23} = -\frac{3 \Xi_{27}^{0}}{\sqrt{30}}, \qquad
\nonumber \\ &&
W^{13}_{11} = \frac{\Xi_{27}^{'-}}{\sqrt{6}} + \frac{2 \Xi_{27}^{-}}{\sqrt{30}}, \qquad
W^{23}_{12} = -\frac{\Xi_{27}^{'-}}{\sqrt{6}} + \frac{\Xi_{27}^{-}}{\sqrt{30}}, \qquad
W^{33}_{13} = -\frac{3 \Xi_{27}^{-}}{\sqrt{30}}, \qquad
W^{23}_{11} = \frac{\Xi_{27}^{'--}}{\sqrt{2}}, \qquad
\end{eqnarray}

\textbullet\ $S = -3, I=1$
\begin{eqnarray}
&&
W^{33}_{22} = \Omega_{27}^{0}, \qquad
W^{33}_{12} = \frac{\Omega_{27}^-}{\sqrt{2}}, \qquad
W^{33}_{11} = \Omega_{27}^{--}.
\end{eqnarray}
where the primed particles are associated with the higher isospin representation in each strangeness multiplet (see Section \ref{su3} for the $27$ weight diagram), i.e. for the $S=-1$ multiplet, we have the triple state: $\Sigma_{27}^{'0}$ has $I=2$, $\Sigma_{27}^{0}$ has $I=1$, and $\Lambda_{27}^{0}$ has $I=0$.

The irreducible tensors associated with the decomposition in Eq. \ref{eq:decomp1} are:
\begin{eqnarray}
8^{i}_{j} &=& [T^8]^{m}_{n}W^{ni}_{mj} \nonumber\\
10^{ijk} &=& \epsilon^{irs}[T^8]^{t}_{r}W^{jk}_{st} + \epsilon^{krs}[T^8]^{t}_{r}W^{ij}_{st} + \epsilon^{jrs}[T^8]^{t}_{r}W^{ki}_{st} \nonumber\\
\overline{10}_{ijk} &=& \epsilon_{irs}[T^8]^{r}_{t}W^{st}_{jk} + \epsilon_{krs}[T^8]^{r}_{t}W^{st}_{ij} + \epsilon_{jrs}[T^8]^{r}_{t}W^{st}_{ki} \nonumber\\
(27_{a})^{ik}_{jl} &=& ([T^8]^{m}_{j}W^{ik}_{ml} + [T^8]^{m}_{l}W^{ik}_{mj}) - \frac{1}{5}(\delta^{i}_{j}[T^8]^{m}_{n}W^{nk}_{ml} + \delta^{i}_{l}[T^8]^{m}_{n}W^{nk}_{mj} + \delta^{k}_{j}[T^8]^{m}_{n}W^{ni}_{ml} + \delta^{k}_{l}[T^8]^{m}_{n}W^{ni}_{mj}) \nonumber\\
(27_{b})^{ik}_{jl} &=& ([T^8]^{i}_{m}W^{mk}_{jl} + [T^8]^{k}_{m}W^{im}_{jl}) - \frac{1}{5}(\delta^{i}_{j}[T^8]^{m}_{n}W^{nk}_{ml} + \delta^{i}_{l}[T^8]^{m}_{n}W^{nk}_{mj} + \delta^{k}_{j}[T^8]^{m}_{n}W^{ni}_{ml} + \delta^{k}_{l}[T^8]^{m}_{n}W^{ni}_{mj}) \nonumber\\
\label{eq:27tensor}
\end{eqnarray}
where we have only given the irreducible representations that are not orthogonal to the representations contained in $8 \otimes 8$. In terms of reduced matrix elements, the decay amplitude then takes the form:
\begin{eqnarray}
\langle \mathbf{8} \otimes \mathbf{8} | H | \mathbf{27} \rangle &=& \alpha_{27} * (27^*)^{jl}_{ik}W^{jl}_{ik} + 
\beta_{27} * {(8_a^*)^{n}_{m}} {8^{n}_{m}} + \gamma_{27} * {(8_b^*)^{n}_{m}} {8^{n}_{m}} + \delta_{27} * (10^*)^{ijk} {10}^{ijk} +  \epsilon_{27} * (\overline{10}^*)_{lmn} {\overline{10}}_{lmn}   \nonumber\\
& &+ \psi_{27} * (27^*)^{mn}_{kl} {(27_a)^{mn}_{kl}} + \zeta_{27} * {(27^*)^{mn}_{kl}} {(27_b)^{mn}_{kl}} \nonumber\\
\label{eq:decay2}
\end{eqnarray}
where as in Eq. \ref{eq:decay1}, the first tensor corresponds to $8 \otimes 8$ with the baryon octet and meson tensor coefficient complex-conjugated and transposed. There are two terms containing octets because the $8$ appears twice in the $8 \otimes 8$ decomposition, and there are two terms containing the $27$ tensors because, as in Eq. \ref{eq:27tensor}, there are two $27$ representations appearing in the $27 \otimes 8$ decomposition of Eq. \ref{eq:decomp1}. The first term in the amplitude corresponds to the singlet piece of the Hamiltonian and as such is the dominant contribution. The subsequent six terms are all from $SU(3)$ breaking and hence are on the order of one-third the strength of $\alpha_{27}$. The coefficients of the reduced matrix elements associated with each decay mode of the $27$ are summarized in Appendix \ref{27table}, with the particle assignments and associated weight diagram in Appendix \ref{su3}.

\section{The $35$ Decay Amplitude}\label{35}
The $35$ exotic multiplet may also decay into an octet baryon + pseudoscalar meson. Again, $SU(3)$ flavor symmetry will allow us to parametrize all the decay modes in terms of just a few parameters. Since the decomposition of the tensor product $8 \otimes 8$ in Eq. \ref{eq:decomp} does not contain a $35$, there will be no singlet contribution from the effective Hamiltonian in Eq. \ref{eq:heff}. The leading order contribution will be from the $SU(3)$ breaking term.

Since the $SU(3)$ breaking term in Eq. \ref{eq:heff} transforms like an $8$, we need the decomposition of $35 \otimes 8$ into irreducible representations in order to find out how many reduced matrix elements there are in the decay amplitude. The decomposition is:
\begin{eqnarray}
35 \otimes 8 &=&  10 \oplus  27 \oplus 28 \oplus 35 \oplus 35 \oplus 64 \oplus 81
\label{eq:decomp2}
\end{eqnarray}
Only the $10$ and $27$ multiplets in this decomposition have non-vanishing matrix elements with $8 \otimes 8$ by referring back to Eq. \ref{eq:decomp}, thus there will only be two reduced matrix elements in the decay amplitude:
\begin{eqnarray}
\langle {\bf8} \otimes {\bf8} | H | {\bf{35}} \rangle &=& \beta_{35} * (10^*)^{lmn} 10^{lmn} + \gamma_{35} * {(27^*)^{mn}_{kl}} {27^{mn}_{kl}}
\label{eq:decay3}
\end{eqnarray}
The $35$ state is:
\begin{equation}
| 35 \rangle = F^{ijkl}_m |^{m}_{ijkl} \rangle
\end{equation}
where tensor $F^{ijkl}_m$ is traceless and symmetric in its upper indices with coefficients:

\textbullet\ $S = 1, I=2$
\begin{eqnarray}
&&
F_3^{1111} = \Theta_{35}^{+++}, \qquad
F_3^{1112} = \frac{\Theta_{35}^{++}}{2}, \qquad
F_3^{1122} = \frac{\Theta_{35}^{+}}{\sqrt{6}}, \qquad
F_3^{1222} = \frac{\Theta_{35}^{0}}{2}, \qquad
F_3^{2222} = \Theta_{35}^{-}, \qquad
\end{eqnarray}

\textbullet\ $S = 0, I=5/2,3/2$
\begin{eqnarray}
&&
F_2^{1111} = \Delta_{35}^{'+++}, \qquad
F_1^{1111} = \frac{\Delta_{35}^{'++}}{\sqrt{5}} + \frac{
2 \Delta_{35}^{++}}{\sqrt{30}}, \qquad
F_2^{1112} = -\frac{\Delta_{35}^{'++}}{\sqrt{5}} + \frac{
\Delta_{35}^{++}}{2 \sqrt{30}}, \qquad
\nonumber \\ &&
F_{3}^{1113} = - \frac{5 \Delta_{35}^{++}}{2 \sqrt{30}}, \qquad
F_1^{1112} = \frac{\Delta_{35}^{'+}}{\sqrt{10}} + \frac{
\Delta_{35}^{+}}{2 \sqrt{10}}, \qquad
F_2^{1122} = -\frac{\Delta_{35}^{'+}}{\sqrt{10}} + \frac{
\Delta_{35}^{+}}{3 \sqrt{10}}, \qquad
\nonumber \\ &&
F_{3}^{1123} = - \frac{5
\Delta_{35}^{+}}{6 \sqrt{10}}, \qquad
F_1^{1122} = \frac{\Delta_{35}^{'0}}{\sqrt{10}} + \frac{
\Delta_{35}^0}{3 \sqrt{10}}, \qquad
F_2^{1222} = -\frac{\Delta_{35}^{'0}}{\sqrt{10}} + \frac{
\Delta_{35}^0}{2 \sqrt{10}}, \qquad
\nonumber \\ &&
F_3^{1223} = - \frac{5
\Delta_{35}^0}{6 \sqrt{10}}, \qquad
F_1^{1222} = \frac{\Delta_{35}^{'-}}{\sqrt{5}} + \frac{
\Delta_{35}^-}{2 \sqrt{30}}, \qquad
F_2^{2222} = -\frac{\Delta_{35}^{'-}}{\sqrt{5}} + \frac{2
\Delta_{35}^-}{\sqrt{30}}, \qquad
\nonumber \\ &&
F_3^{2223} = -\frac{5 \Delta_{35}^-}{2 \sqrt{30}}, \qquad
F_1^{2222} =  \Delta_{35}^{'--}
\end{eqnarray}

\textbullet\ $S = -1, I=2,1$
\begin{eqnarray}
&&
F_2^{1113} = \frac{\Sigma_{35}^{'++}}{2}, \qquad
F_1^{1113} = \frac{\Sigma_{35}^{'+}}{4} + \frac{
\Sigma_{35}^+}{4}, \qquad
F_2^{1123} = -\frac{\Sigma_{35}^{'+}}{4} + \frac{
\Sigma_{35}^+}{12}, \qquad
F_3^{1133} = -\frac{
\Sigma_{35}^+}{3},
\nonumber \\ &&
F_1^{1123} = \frac{\Sigma_{35}^{'0}}{2 \sqrt{6}} + \frac{\Sigma_{35}^0}{6 \sqrt{2}}, \qquad
F_2^{1223} = -\frac{\Sigma_{35}^{'0}}{2 \sqrt{6}} + \frac{\Sigma_{35}^0}{6 \sqrt{2}}, \qquad
F_3^{1233} = -\frac{\Sigma_{35}^0}{3 \sqrt{2}}, \qquad
\nonumber \\ &&
F_1^{1223} = \frac{\Sigma_{35}^{'-}}{4} + \frac{
\Sigma_{35}^-}{12}, \qquad
F_2^{2223} = -\frac{\Sigma_{35}^{'-}}{4} + \frac{
\Sigma_{35}^-}{4}, \qquad
F_3^{2233} = - \frac{
\Sigma_{35}^-}{3}, \qquad
F_1^{2223} = \frac{\Sigma_{35}^{'--}}{2}.
\end{eqnarray}

\textbullet\ $S = -2, I=3/2,1/2$
\begin{eqnarray}
&&
F_2^{1133} = \frac{\Xi_{35}^{'+}}{\sqrt{6}}, \qquad
F_1^{1133} = \frac{\Xi_{35}^{'0}}{3 \sqrt{2}} + \frac{\Xi_{35}^0}{3 \sqrt{2}},
\qquad
F_2^{1233} = -\frac{\Xi_{35}^{'0}}{3 \sqrt{2}} + \frac{\Xi_{35}^0}{6 \sqrt{2}},
\qquad
F_3^{1333} = - \frac{\Xi_{35}^0}{2 \sqrt{2}},
\nonumber \\ &&
F_1^{1233} = \frac{\Xi_{35}^{'-}}{3 \sqrt{2}} + \frac{\Xi_{35}^-}{6 \sqrt{2}},
\qquad
F_2^{2233} = -\frac{\Xi_{35}^{'-}}{3 \sqrt{2}} + \frac{\Xi_{35}^-}{3 \sqrt{2}},
\qquad
F_3^{2333} = - \frac{\Xi_{35}^-}{2 \sqrt{2}}, \qquad
F_1^{2233} = \frac{\Xi_{35}^{'--}}{\sqrt6}
\end{eqnarray}

\textbullet\ $S = -3, I=1,0$
\begin{eqnarray}
&&
F_2^{1333} = \frac{\Omega_{35}^{'0}}{2}, \qquad
F_1^{1333} = \frac{\Omega_{35}^{'-}}{2 \sqrt{2}} + \frac{\Omega_{35}^{-}}{2 \sqrt{3}}, \qquad
F_2^{2333} = -\frac{\Omega_{35}^{'-}}{2 \sqrt{2}} + \frac{\Omega_{35}^{-}}{2 \sqrt{3}}, \qquad
\nonumber \\ &&
F_3^{3333} = -\frac{\Omega_{35}^{-}}{\sqrt{3}}, \qquad
F_1^{2333} = \frac{\Omega_{35}^{'--}}{2}.
\end{eqnarray}

\textbullet\ $S = -4, I=1/2$
\begin{eqnarray}
F_2^{3333} = \Phi^-, \qquad
F_1^{3333} = \Phi^{--}.
\end{eqnarray}
where the primed greek letters correspond to the higher isospin multiplet for a given strangeness, i.e. for $S=-1$, $\Sigma_{35}^{'0}$ has $I=2$, and $\Sigma_{35}^{0}$ has $I=1$. The $SU(3)$ weight diagram for the $35$ is shown in Appendix \ref{su3}.

The irreducible tensors associated with the decomposition of Eq. \ref{eq:decomp2} are:
\begin{eqnarray}
10^{ijk} &=& [T^8]^{m}_{n} F^{nijk}_{m} \nonumber\\
27^{ik}_{jl} &=& \epsilon_{jst}[T^8]^{s}_{m}F^{tmik}_{l} +  \epsilon_{lst}[T^8]^{s}_{m}F^{tmik}_{j}
\nonumber\\
\label{eq:35x8}
\end{eqnarray}
We may now evaluate the decay amplitude, Eq. \ref{eq:decay3}, using the irreducible tensors for $8 \otimes 8$ from Eq. \ref{eq:8x8} and those of Eq. \ref{eq:35x8}. The coefficients of the reduced matrix elements for each decay mode of the $35$ are tabulated in Appendix \ref{35table}.

\section{Discussion}
The decay amplitudes for all isomultiplets in the $\overline{10}$, $27$, and $35$ can now be computed by reading off the coefficients corresponding to a particular decay mode tabulated in appendices \ref{table1}, \ref{27table}, and \ref{35table}. One may then calculate the decay width with appropriate spin-averaging factors as:
\begin{equation}
\frac{d\Gamma}{d\Omega} = \frac{|a_{if}|^2}{32\pi^2}\frac{p}{M^2} = \frac{|a_{if}|^2}{64\pi^2}\frac{[M^4 - 2({m_1}^2 + {m_2}^2)M^2 + ({m_1}^2 - {m_2}^2)^2]^{1/2}}{M^3}
\end{equation}
where $a_{if}$ is the decay amplitude, and the two-body phase space factor has been made explicit.

Current experimental data shows a range of values for the width of $\Theta_{\overline{10}}$, and in most cases it falls below the detector resolution. Improvements in the measurement of the width of $\Theta_{\overline{10}}$ as well as measurements of the widths of additional pentaquarks are necessary in order for the decay amplitudes, and hence decay widths and branching ratios, to be determined for all members of the $SU(3)$ multiplet. However, if enough widths are measured, and one is able to obtain values for all the reduced matrix elements corresponding to a given multiplet, then  it will be possible to calculate the partial decay widths for all members of the multiplet. One may then compare these results to the various theoretical models and potentially rule some of these out.

\section{Acknowledgements}
The authors would like to thank Benjamin Grinstein for many engaging discussions and suggestions. SMG would like to thank Elizabeth Jenkins for her invaluable guidance in understanding the group theoretical aspects of pentaquarks. This work was supported in part by the Department of Energy under Grant DE-FG03-97ER40546.

\appendix
\section{$SU(3)$ Weight Diagrams}
\label{su3}
\begin{picture}(150,100)(-140,0)
\put(55,60){ $\Theta_{\bf \overline{10}}${\small (0,+1)}}
\put(55,40){ $N_{\bf \overline{10}}${\small ($\frac 1 2$,\,0)}}
\put(55,20){ $\Sigma_{\bf \overline{10}}${\small (1,\,-1)}}
\put(55,0){ $\Xi^\prime_{\bf \overline{10}}${\small ($\frac 3 2$,\,-2)}}
\put(0,60){\circle*{3}}
\put(-10,40){\circle*{3}}
\put(10,40){\circle*{3}}
\put(-20,20){\circle*{3}}
\put(0,20){\circle*{3}}
\put(20,20){\circle*{3}}
\put(-30,0){\circle*{3}}
\put(-10,0){\circle*{3}}
\put(10,0){\circle*{3}}
\put(30,0){\circle*{3}}
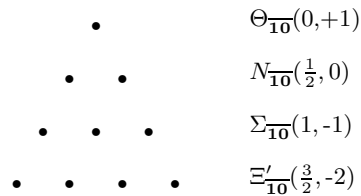
\put(-50,-40){ {\rm \scriptsize{{Figure 1}:  $SU(3)$ weight diagram of
$\overline{{\bf 10}}$. Labels $(I,S)$ }}} 
\put(-13,-50){{\rm \scriptsize{to the right of a particle isomultiplet stand}}}
\put(-13,-60){{\rm \scriptsize{for isospin and strangeness.}}}
\end{picture}
\vspace{2cm}

\begin{picture}(130,100)(-140,0)
\put(65,60){ $\Theta_{\bf 27}${\small (1,+1)}}
\put(110,40){ $\Delta_{\bf 27}${\small ($\frac 3 2$,\,0)}}
\put(65,40){ $N_{\bf 27}${\small ($\frac 1 2$,\,0)},}
\put(110,20){ $\Sigma_{\bf 27}${\small (1,\,-1)},}
\put(155,20){ $\Sigma^\prime_{\bf 27}${\small (2,\,-1)}}
\put(65,20){ $\Lambda_{\bf 27}${\small (0,\,-1)},}
\put(65,0){ $\Xi_{\bf 27}${\small ($\frac 1 2$,\,-2)},}
\put(110,0){ $\Xi^\prime_{\bf 27}${\small ($\frac 3 2$,\, -2)}}
\put(65,-20){ $\Omega^\prime_{\bf 27}${\small (1,\,-3)}}
\put(-20,60){\circle*{3}}
\put(0,60){\circle*{3}}
\put(20,60){\circle*{3}}
\put(-30,40){\circle*{3}}
\put(-10,40){\circle{6}}
\put(-10,40){\circle*{3}}
\put(10,40){\circle{6}}
\put(10,40){\circle*{3}}
\put(30,40){\circle*{3}}
\put(-40,20){\circle*{3}}
\put(-20,20){\circle{6}}
\put(-20,20){\circle*{3}}
\put(0,20){\circle{10}}
\put(0,20){\circle{6}}
\put(0,20){\circle*{3}}
\put(20,20){\circle{6}}
\put(20,20){\circle*{3}}
\put(40,20){\circle*{3}}
\put(-30,0){\circle*{3}}
\put(-10,0){\circle{6}}
\put(-10,0){\circle*{3}}
\put(10,0){\circle{6}}
\put(10,0){\circle*{3}}
\put(30,0){\circle*{3}}
\put(-20,-20){\circle*{3}}
\put(0,-20){\circle*{3}}
\put(20,-20){\circle*{3}}
\end{picture}
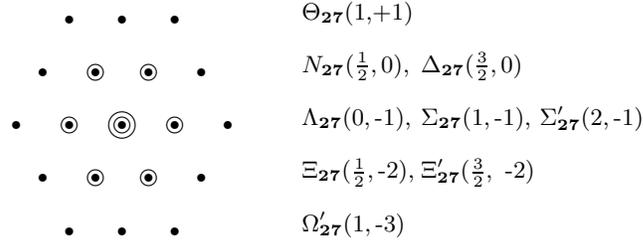
\put(-45,-40){ {\rm \scriptsize{{Figure 2}:  $SU(3)$ weight diagram of  ${\bf 27}$.}}}

\vspace{1cm}

\begin{picture}(100,100)(-140,0)
\put(75,60){ $\Theta_{\bf 35}${\small (2,+1)}}
\put(75,40){ $\Delta_{\bf 35}${\small ($\frac 3 2$,\,0)},\ $\Delta^\prime_{\bf 35}${\small ($\frac 5 2$,\,0)}}
\put(75,20){ $\Sigma_{\bf 35}${\small (1,\,-1)},\ $\Sigma^\prime_{\bf 35}${\small (1,\,-1)}}
\put(75,0){ $\Xi_{\bf 35}${\small ($\frac 1 2$,\,-2)},\ $\Xi^\prime_{\bf 35}${\small ($\frac 3 2$,\,-2)}}
\put(75,-20){ $\Omega_{\bf 35}${\small (0,\,-3)},\ $\Omega^\prime_{\bf 35}${\small (1,\,-3)}}
\put(75,-40){ $\Phi_{\bf 35}${\small ($\frac 1 2$,\,-4)}}
\put(-40,60){\circle*{3}}
\put(-20,60){\circle*{3}}
\put(-0,60){\circle*{3}}
\put(20,60){\circle*{3}}
\put(40,60){\circle*{3}}
\put(-50,40){\circle*{3}}
\put(-30,40){\circle{6}}
\put(-30,40){\circle*{3}}
\put(-10,40){\circle{6}}
\put(-10,40){\circle*{3}}
\put(10,40){\circle{6}}
\put(10,40){\circle*{3}}
\put(30,40){\circle{6}}
\put(30,40){\circle*{3}}
\put(50,40){\circle*{3}}
\put(-40,20){\circle*{3}}
\put(-20,20){\circle{6}}
\put(-20,20){\circle*{3}}
\put(-0,20){\circle{6}}
\put(0,20){\circle*{3}}
\put(20,20){\circle{6}}
\put(20,20){\circle*{3}}
\put(40,20){\circle*{3}}
\put(-30,0){\circle*{3}}
\put(-10,0){\circle{6}}
\put(-10,0){\circle*{3}}
\put(10,0){\circle{6}}
\put(10,0){\circle*{3}}
\put(30,0){\circle*{3}}
\put(-20,-20){\circle*{3}}
\put(0,-20){\circle{6}}
\put(0,-20){\circle*{3}}
\put(20,-20){\circle*{3}}
\put(-10,-40){\circle*{3}}
\put(10,-40){\circle*{3}}
\end{picture}
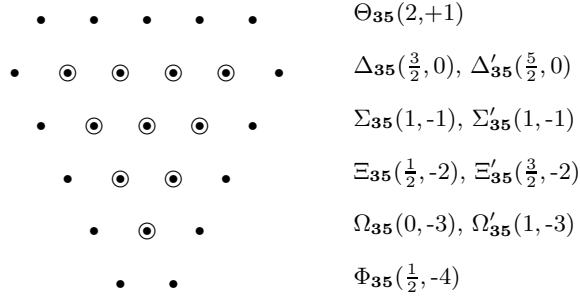
\put(-13,-60){ {\rm \scriptsize{{Figure 3}:  $SU(3)$ weight diagram of   
${\bf 35}$.}}}

\begin{section}{Decay Amplitudes of $\bar{10}$ Pentaquark Multiplet}
\label{table1}
\begin{center}
\begin{tabular}{|c|c|c|c|c|c|} \hline
$\Theta_{\overline{10}}^+$ & $\alpha_{\overline{10}}$ & $\beta_{\overline{10}}$ & $\gamma_{\overline{10}}$ & $\delta_{\overline{10}}$ & $\epsilon_{\overline{10}}$ \\ \hline
$PK^0$ & $1$ & $-2\sqrt{3}$ & 0 & 0 & 0 \\ \hline
$NK^+$ & $-1$ & $2\sqrt{3}$ & 0 & 0 & 0 \\ \hline
\end{tabular}
\end{center}

\begin{center}
\begin{tabular}{|c|c|c|c|c|c|} \hline
$N^+_{\bar{10}}$ & $\alpha_{\overline{10}}$ & $\beta_{\overline{10}}$ & $\gamma_{\overline{10}}$ & $\delta_{\overline{10}}$ & $\epsilon_{\overline{10}}$ \\ \hline
$P\pi^0$ & $-\sqrt{1/6}$ & $\sqrt{1/2}$ & 0 & $\sqrt{1/2}$ & $-1/5\sqrt{6}$ \\ \hline
$N\pi^+$ & $-\sqrt{1/3}$ & $1$ & 0 & $1$ & $-1/5\sqrt{3}$ \\ \hline
$P\eta$ & $\sqrt{1/2}$ & $-\sqrt{3/2}$ & $-\sqrt{2/3}$ & $\sqrt{1/6}$ & $-3/5\sqrt{2}$ \\ \hline
$\Lambda K^+$ & $-\sqrt{1/2}$ & $\sqrt{3/2}$ & $\sqrt{1/6}$ & $-\sqrt{2/3}$ & $-3/5\sqrt{2}$ \\ \hline
$\Sigma^0K^+$ & $\sqrt{1/6}$ & $-\sqrt{1/2}$ & $\sqrt{1/2}$ & $0$ & $-1/5\sqrt{6}$ \\ \hline
$\Sigma^+K^0$ & $\sqrt{1/3}$ & $-1$ & $1$ & 0 & $-1/5\sqrt{3}$ \\ \hline
\end{tabular}
\end{center}

\begin{center}
\begin{tabular}{|c|c|c|c|c|c|} \hline
$N^0_{\bar{10}}$ & $\alpha_{\overline{10}}$ & $\beta_{\overline{10}}$ & $\gamma_{\overline{10}}$ & $\delta_{\overline{10}}$ & $\epsilon_{\overline{10}}$ \\ \hline
$N\pi^0$ & $-\sqrt{1/6}$ & $\sqrt{1/2}$ & 0 & $\sqrt{1/2}$ & $-1/5\sqrt{6}$ \\ \hline
$P\pi^-$ & $\sqrt{1/3}$ & $-1$ & 0 & $-1$ & $1/5\sqrt{3}$ \\ \hline
$N\eta$ & $-\sqrt{1/2}$ & $\sqrt{3/2}$ & $\sqrt{2/3}$ & $-\sqrt{1/6}$ & $3/5\sqrt{2}$ \\ \hline
$\Lambda K^0$ & $\sqrt{1/2}$ & $-\sqrt{3/2}$ & $-\sqrt{1/6}$ & $\sqrt{2/3}$ & $3/5\sqrt{2}$ \\ \hline
$\Sigma^0K^0$ & $\sqrt{1/6}$ & $-\sqrt{1/2}$ & $\sqrt{1/2}$ & 0 & $-1/5\sqrt{6}$ \\ \hline
$\Sigma^-K^+$ & $-\sqrt{1/3}$ & $1$ & $-1$ & 0 &$1/5\sqrt{3}$ \\ \hline
\end{tabular}
\end{center}

\begin{center}
\begin{tabular}{|c|c|c|c|c|c|} \hline
$\Sigma_{\overline{10}}^+$ & $\alpha_{\overline{10}}$ & $\beta_{\overline{10}}$ & $\gamma_{\overline{10}}$ & $\delta_{\overline{10}}$ & $\epsilon_{\overline{10}}$ \\ \hline
$P\overline{K^0}$ & $-\sqrt{1/3}$ & 0 & 1 & 0 & $4/5\sqrt{3}$ \\ \hline
$\Xi^0K^+$ & $\sqrt{1/3}$ & 0 & 0 & 1 & $4/5\sqrt{3}$ \\ \hline
$\Lambda\pi^+$ & $-\sqrt{1/2}$ & 0 & $\sqrt{1/6}$ & $\sqrt{1/6}$ & $-2\sqrt{2}/5$ \\ \hline
$\Sigma^+\eta$ & $\sqrt{1/2}$ & 0 & $\sqrt{1/6}$ & $\sqrt{1/6}$ & $-2\sqrt{2}/5$ \\ \hline 
$\Sigma^0\pi^+$ & $\sqrt{1/6}$ & 0 & $\sqrt{1/2}$ & $-\sqrt{1/2}$ & 0 \\ \hline
$\Sigma^+\pi^0$ & $-\sqrt{1/6}$ & 0 & $-\sqrt{1/2}$ & $\sqrt{1/2}$ & 0 \\ \hline
\end{tabular}
\end{center}

\begin{center}
\begin{tabular}{|c|c|c|c|c|c|} \hline
$\Sigma_{\overline{10}}^0$ & $\alpha_{\overline{10}}$ & $\beta_{\overline{10}}$ & $\gamma_{\overline{10}}$ & $\delta_{\overline{10}}$ & $\epsilon_{\overline{10}}$ \\ \hline
$\Lambda\pi^0$ & $-\sqrt{1/2}$ & 0 & $\sqrt{1/6}$ & $\sqrt{1/6}$ & $-2\sqrt{2}/5$  \\ \hline
$\Sigma^0\eta$ & $\sqrt{1/2}$ & 0 & $\sqrt{1/6}$ & $\sqrt{1/6}$ & $-2\sqrt{2}/5$ \\ \hline
$\Xi^0K^0$ & $-\sqrt{1/6}$ & 0 & 0 & $-\sqrt{1/2}$ & $-2\sqrt{2}/5\sqrt{3}$ \\ \hline
$\Xi^-K^+$ & $\sqrt{1/6}$ & 0 & 0 & $\sqrt{1/2}$ & $2\sqrt{2}/5\sqrt{3}$ \\ \hline
$N\overline{K^0}$ & $\sqrt{1/6}$ & 0 & $-\sqrt{1/2}$ & 0 & $-2\sqrt{2}/5\sqrt{3}$ \\ \hline 
$PK^-$ & $-\sqrt{1/6}$ & 0 & $\sqrt{1/2}$ & 0 & $2\sqrt{2}/5\sqrt{3}$ \\ \hline
$\Sigma^-\pi^+$ & $-\sqrt{1/6}$ & 0 & $-\sqrt{1/2}$ & $\sqrt{1/2}$ & 0 \\ \hline
$\Sigma^+\pi^-$ & $\sqrt{1/6}$ & 0 & $\sqrt{1/2}$ & $-\sqrt{1/2}$ & 0 \\ \hline
\end{tabular}
\end{center}

\begin{center}
\begin{tabular}{|c|c|c|c|c|c|} \hline
$\Sigma_{\overline{10}}^-$ &  $\alpha_{\overline{10}}$ & $\beta_{\overline{10}}$ & $\gamma_{\overline{10}}$ & $\delta_{\overline{10}}$ & $\epsilon_{\overline{10}}$\\ \hline
$NK^-$ & $\sqrt{1/3}$ & 0 & $-1$ & 0 & $-4/5\sqrt{3}$ \\ \hline
$\Xi^-K^0$ & $-\sqrt{1/3}$ & 0 & 0 & $-1$ & $-4/5\sqrt{3}$ \\ \hline
$\Lambda\pi^-$ & $\sqrt{1/2}$ & 0 & $-\sqrt{1/6}$ & $-\sqrt{1/6}$ & $2\sqrt{2}/5$ \\ \hline
$\Sigma^-\eta$ & $-\sqrt{1/2}$ & 0 & $-\sqrt{1/6}$ & $-\sqrt{1/6}$ & $2\sqrt{2}/5$ \\ \hline
$\Sigma^0\pi^-$ & $\sqrt{1/6}$ & 0 & $\sqrt{1/2}$ & $-\sqrt{1/2}$ & 0 \\ \hline
$\Sigma^-\pi^0$ & $-\sqrt{1/6}$ & 0 & $-\sqrt{1/2}$ & $\sqrt{1/2}$ & 0\\ \hline
\end{tabular}
\end{center}

\begin{center}
\begin{tabular}{|c|c|c|c|c|c|} \hline
$\Xi_{\overline{10}}^{+}$ & $\alpha_{\overline{10}}$ & $\beta_{\overline{10}}$ & $\gamma_{\overline{10}}$ & $\delta_{\overline{10}}$ & $\epsilon_{\overline{10}}$ \\ \hline
$\Xi^0\pi^+$ & 1 & $\sqrt{3}$ & 0 & 0 & 1 \\ \hline
$\Sigma^+\overline{K^0}$ & -1 & -$\sqrt{3}$ & 0 & 0 &1 \\ \hline
\end{tabular}
\end{center}

\begin{center}
\begin{tabular}{|c|c|c|c|c|c|} \hline
$\Xi_{\overline{10}}^0$ & $\alpha_{\overline{10}}$ & $\beta_{\overline{10}}$ & $\gamma_{\overline{10}}$ & $\delta_{\overline{10}}$ & $\epsilon_{\overline{10}}$\\ \hline
$\Sigma^0\overline{K^0}$ & $-\sqrt{2/3}$ & $-\sqrt{2}$ & 0 & 0 & $\sqrt{2/3}$ \\ \hline
$\Sigma^+K^-$ & $-\sqrt{1/3}$ & $-1$ & 0 & 0 & $\sqrt{1/3}$ \\ \hline
$\Xi^0\pi^0$ & $\sqrt{2/3}$ & $\sqrt{2}$ & 0 & 0 &$\sqrt{2/3}$ \\ \hline
$\Xi^-\pi^+$ & $\sqrt{1/3}$ & $1$ & 0 & 0 & $\sqrt{1/3}$ \\  \hline
\end{tabular}
\end{center}

\begin{center}
\begin{tabular}{|c|c|c|c|c|c|} \hline
$\Xi_{\overline{10}}^-$ & $\alpha_{\overline{10}}$ & $\beta_{\overline{10}}$ & $\gamma_{\overline{10}}$ & $\delta_{\overline{10}}$ & $\epsilon_{\overline{10}}$ \\ \hline
$\Sigma^0K^-$ & $-\sqrt{2/3}$ & $-\sqrt{2}$ & 0 & 0 & $\sqrt{2/3}$ \\ \hline
$\Sigma^-\overline{K^0}$ & $\sqrt{1/3}$ & $1$ & 0 & 0 & $-\sqrt{1/3}$ \\ \hline
$\Xi^0\pi^-$ & $-\sqrt{1/3}$ & $-1$ & 0 &0 & $-\sqrt{1/3}$ \\ \hline
$\Xi^-\pi^0$ & $\sqrt{2/3}$ & $\sqrt{2}$ & 0 & 0 & $\sqrt{2/3}$  \\ \hline
\end{tabular}
\end{center}

\begin{center}
\begin{tabular}{|c|c|c|c|c|c|} \hline
$\Xi_{\overline{10}}^{--}$ & $\alpha_{\overline{10}}$ & $\beta_{\overline{10}}$ & $\gamma_{\overline{10}}$ & $\delta_{\overline{10}}$ & $\epsilon_{\overline{10}}$ \\ \hline
$\Xi^-\pi^-$ & -1 & -$\sqrt{3}$ & 0 & 0 & -1 \\ \hline
$\Sigma^-K^-$ & 1 & $\sqrt{3}$  & 0 & 0 & -1 \\ \hline
\end{tabular}
\end{center}
\end{section}

\begin{section}{Decay Amplitudes of $27$ Pentaquark Multiplet}
\label{27table}
\begin{center}
\begin{tabular}{|c|c|c|c|c|c|c|c|} \hline
$\Theta^{++}_{27}$ & $\alpha_{27}$ & $\beta_{27}$ & $\gamma_{27}$ & $\delta_{27}$ & $\epsilon_{27}$ & $\zeta_{27}$ & $\psi_{27}$\\ \hline
$PK^+$ & $\sqrt{2}$ & $0$ & $0$ & $0$ & 0 & $-8$ & $4$ \\ \hline
\end{tabular}
\end{center}

\begin{center}
\begin{tabular}{|c|c|c|c|c|c|c|c|} \hline
$\Theta^+_{27}$ & $\alpha_{27}$ & $\beta_{27}$ & $\gamma_{27}$ & $\delta_{27}$ & $\epsilon_{27}$ & $\zeta_{27}$ & $\psi_{27}$\\ \hline
$PK^0$ & $1$ & $0$ & $0$ & $0$ & 0 & $-4\sqrt{2}$ & $2\sqrt{2}$ \\ \hline
$NK^+$ & $1$ & $0$ & $0$ & $0$ & 0 & $-4\sqrt{2}$ & $2\sqrt{2}$ \\ \hline
\end{tabular}
\end{center}

\begin{center}
\begin{tabular}{|c|c|c|c|c|c|c|c|} \hline
$\Theta^0_{27}$ & $\alpha_{27}$ & $\beta_{27}$ & $\gamma_{27}$ & $\delta_{27}$ & $\epsilon_{27}$ & $\zeta_{27}$ & $\psi_{27}$\\ \hline
$NK^0$ & $\sqrt{2}$ & $0$ & $0$ & $0$ & 0 & $-8$ & $4$ \\ \hline
\end{tabular}
\end{center}

\begin{center}
\begin{tabular}{|c|c|c|c|c|c|c|c|} \hline
$N^+_{27}$ & $\alpha_{27}$ & $\beta_{27}$ & $\gamma_{27}$ & $\delta_{27}$ & $\epsilon_{27}$ & $\zeta_{27}$ & $\psi_{27}$\\ \hline
$P\pi^0$ & $\sqrt{1/30}$ & $0$ & $3/2$ & $0$ & $-\sqrt{3/5}$ & $-14/5\sqrt{15}$ & $1/5\sqrt{15}$ \\ \hline
$N\pi^+$ & $\sqrt{1/15}$ & $0$ & $3/\sqrt{2}$ & $0$ & $-\sqrt{6/5}$ & $-14\sqrt{2}/5\sqrt{15}$ & $\sqrt{2}/5\sqrt{15}$ \\ \hline
$P\eta$ & $3/\sqrt{10}$ & $-\sqrt{3}$ & $\sqrt{3}/2$ & $0$ & $3/\sqrt{5}$ & $-42/5\sqrt{5}$ & $3/5\sqrt{5}$ \\ \hline
$\Lambda K^+$ & $3/\sqrt{10}$ & $\sqrt{3}/2$ & $-\sqrt{3}$ & $0$ & $-3/\sqrt{5}$ & $-42/5\sqrt{5}$ & $3/5\sqrt{5}$ \\ \hline
$\Sigma^0 K^+$ & $\sqrt{1/30}$ & $3/2$ & $0$ & $0$ & $\sqrt{3/5}$ & $-14/5\sqrt{15}$ & $1/5\sqrt{15}$ \\ \hline
$\Sigma^+K^0$ & $\sqrt{1/15}$ & $3/\sqrt{2}$ & $0$ & $0$ & $\sqrt{6/5}$ & $-14\sqrt{2}/5\sqrt{15}$ & $\sqrt{2}/5\sqrt{15}$ \\ \hline
\end{tabular}
\end{center}

\begin{center}
\begin{tabular}{|c|c|c|c|c|c|c|c|} \hline
$N^0_{27}$ & $\alpha_{27}$ & $\beta_{27}$ & $\gamma_{27}$ & $\delta_{27}$ & $\epsilon_{27}$ & $\zeta_{27}$ & $\psi_{27}$\\ \hline
$N\pi^0$ & $-\sqrt{1/30}$ & 0 & $-3/2$ & $0$ & $\sqrt{3/5}$ & $14/5\sqrt{15}$ & $-1/5\sqrt{15}$ \\ \hline
$P\pi^-$ & $\sqrt{1/15}$ & $0$ & $3/\sqrt{2}$ & 0 & $-\sqrt{6/5}$ & $-14\sqrt{2}/5\sqrt{15}$ & $\sqrt{2}/5\sqrt{15}$ \\ \hline
$N\eta$ & $3/\sqrt{10}$ & $-\sqrt{3}$ & $\sqrt{3}/2$ & 0 & $3/\sqrt{5}$ & $-42/5\sqrt{5}$ & $3/5\sqrt{5}$ \\ \hline
$\Lambda K^0$ & $3/\sqrt{10}$ & $\sqrt{3}/2$ & $-\sqrt{3}$ & 0 & $-3/\sqrt{5}$ & $-42/5\sqrt{5}$ & $3/5\sqrt{5}$ \\ \hline
$\Sigma^0 K^0$ & $-\sqrt{1/30}$ & $-3/2$ & $0$ & $0$ & $-\sqrt{3/5}$ & $14/5\sqrt{15}$ & $-1/5\sqrt{15}$ \\ \hline
$\Sigma^- K^+$ & $\sqrt{1/15}$ & $3/\sqrt{2}$ & 0 & $0$ & $\sqrt{6/5}$ & $-14\sqrt{2}/5\sqrt{15}$ & $\sqrt{2}/5\sqrt{15}$ \\ \hline
\end{tabular}
\end{center}

\begin{center}
\begin{tabular}{|c|c|c|c|c|c|c|c|} \hline
$\Delta^{++}_{27}$ & $\alpha_{27}$ & $\beta_{27}$ & $\gamma_{27}$ & $\delta_{27}$ & $\epsilon_{27}$ & $\zeta_{27}$ & $\psi_{27}$\\ \hline
$P\pi^+$ & $1$ & $0$ & $0$ & $-3\sqrt{2}$ & $0$ & $-\sqrt{2}$ & $2\sqrt{2}$ \\ \hline
$\Sigma^+ K^+$ & $1$ & $0$ & $0$ & $3\sqrt{2}$ & $0$ & $-\sqrt{2}$ & $2\sqrt{2}$ \\ \hline
\end{tabular}
\end{center}

\begin{center}
\begin{tabular}{|c|c|c|c|c|c|c|c|} \hline
$\Delta^{+}_{27}$ & $\alpha_{27}$ & $\beta_{27}$ & $\gamma_{27}$ & $\delta_{27}$ & $\epsilon_{27}$ & $\zeta_{27}$ & $\psi_{27}$\\ \hline
$P\pi^0$ & $\sqrt{2/3}$ & $0$ & $0$ & $-2\sqrt{3}$ & $0$ & $-2/\sqrt{3}$ & $4/\sqrt{3}$ \\ \hline
$N\pi^+$ & $-\sqrt{1/3}$ & $0$ & $0$ & $\sqrt{6}$ & $0$ & $\sqrt{2/3}$ & $-2\sqrt{2/3}$ \\ \hline
$\Sigma^0K^+$ & $\sqrt{2/3}$ & $0$ & $0$ & $2\sqrt{3}$ & $0$ & $-2/\sqrt{3}$ & $4/\sqrt{3}$ \\ \hline
$\Sigma^+K^0$ & $-\sqrt{1/3}$ & $0$ & $0$ & $-\sqrt{6}$ & $0$ & $\sqrt{2/3}$ & $-2\sqrt{2/3}$ \\ \hline
\end{tabular}
\end{center}

\begin{center}
\begin{tabular}{|c|c|c|c|c|c|c|c|} \hline
$\Delta^{0}_{27}$ & $\alpha_{27}$ & $\beta_{27}$ & $\gamma_{27}$ & $\delta_{27}$ & $\epsilon_{27}$ & $\zeta_{27}$ & $\psi_{27}$\\ \hline
$P\pi^-$ & $-\sqrt{1/3}$ & $0$ & $0$ & $\sqrt{6}$ & $0$ & $\sqrt{2/3}$ & $-2\sqrt{2/3}$ \\ \hline
$N\pi^0$ & $-\sqrt{2/3}$ & $0$ & $0$ & $2\sqrt{3}$ & $0$ & $2/\sqrt{3}$ & $-4/\sqrt{3}$ \\ \hline
$\Sigma^0K^0$ & $-\sqrt{2/3}$ & $0$ & $0$ & $-2\sqrt{3}$ & $0$ & $2/\sqrt{3}$ & $-4/\sqrt{3}$ \\ \hline
$\Sigma^-K^+$ & $-\sqrt{1/3}$ & $0$ & $0$ & $-\sqrt{6}$ & $0$ & $\sqrt{2/3}$ & $-2\sqrt{2/3}$ \\ \hline
\end{tabular}
\end{center}

\begin{center}
\begin{tabular}{|c|c|c|c|c|c|c|c|} \hline
$\Delta^{-}_{27}$ & $\alpha_{27}$ & $\beta_{27}$ & $\gamma_{27}$ & $\delta_{27}$ & $\epsilon_{27}$ & $\zeta_{27}$ & $\psi_{27}$\\ \hline
$N\pi^-$ & $1$ & $0$ & $0$ & $-3\sqrt{2}$ & $0$ & $-\sqrt{2}$ & $2\sqrt{2}$ \\ \hline
$\Sigma^- K^0$ & $1$ & $0$ & $0$ & $3\sqrt{2}$ & $0$ & $-\sqrt{2}$ & $2\sqrt{2}$ \\ \hline
\end{tabular}
\end{center}

\begin{center}
\begin{tabular}{|c|c|c|c|c|c|c|c|} \hline
$\Sigma^{'++}_{27}$ & $\alpha_{27}$ & $\beta_{27}$ & $\gamma_{27}$ & $\delta_{27}$ & $\epsilon_{27}$ & $\zeta_{27}$ & $\psi_{27}$\\ \hline
$\Sigma^+\pi^+$ & $\sqrt{2}$ & $0$ & $0$ & $0$ & 0 & $4$ & $4$ \\ \hline
\end{tabular}
\end{center}

\begin{center}
\begin{tabular}{|c|c|c|c|c|c|c|c|} \hline
$\Sigma^{'+}_{27}$ & $\alpha_{27}$ & $\beta_{27}$ & $\gamma_{27}$ & $\delta_{27}$ & $\epsilon_{27}$ & $\zeta_{27}$ & $\psi_{27}$\\ \hline
$\Sigma^0\pi^+$ & $1$ & $0$ & $0$ & 0 & $0$ & $2\sqrt{2}$ & $2\sqrt{2}$ \\ \hline
$\Sigma^+\pi^0$ & $1$ & $0$ & $0$ & 0 & $0$ & $2\sqrt{2}$ & $2\sqrt{2}$ \\ \hline
\end{tabular}
\end{center}

\begin{center}
\begin{tabular}{|c|c|c|c|c|c|c|c|} \hline
$\Sigma^{'0}_{27}$ & $\alpha_{27}$ & $\beta_{27}$ & $\gamma_{27}$ & $\delta_{27}$ & $\epsilon_{27}$ & $\zeta_{27}$ & $\psi_{27}$\\ \hline
$\Sigma^0\pi^0$ & $2/\sqrt{3}$ & $0$ & $0$ & $0$ & $0$ & $4\sqrt{2/3}$ & $4\sqrt{2/3}$ \\ \hline
$\Sigma^-\pi^+$ & $-\sqrt{1/3}$ & $0$ & $0$ & $0$ & $0$ & $-2\sqrt{2/3}$ & $-2\sqrt{2/3}$ \\ \hline
$\Sigma^+\pi^-$ & $-\sqrt{1/3}$ & $0$ & $0$ & $0$ & $0$ & $-2\sqrt{2/3}$ & $-2\sqrt{2/3}$ \\ \hline
\end{tabular}
\end{center}

\begin{center}
\begin{tabular}{|c|c|c|c|c|c|c|c|} \hline
$\Sigma^{'-}_{27}$ & $\alpha_{27}$ & $\beta_{27}$ & $\gamma_{27}$ & $\delta_{27}$ & $\epsilon_{27}$ & $\zeta_{27}$ & $\psi_{27}$\\ \hline
$\Sigma^0\pi^-$ & $1$ & $0$ & $0$ & 0 & $0$ & $2\sqrt{2}$ & $2\sqrt{2}$ \\ \hline
$\Sigma^-\pi^0$ & $1$ & $0$ & $0$ & 0 & $0$ & $2\sqrt{2}$ & $2\sqrt{2}$ \\ \hline
\end{tabular}
\end{center}

\begin{center}
\begin{tabular}{|c|c|c|c|c|c|c|c|} \hline
$\Sigma^{'--}_{27}$ & $\alpha_{27}$ & $\beta_{27}$ & $\gamma_{27}$ & $\delta_{27}$ & $\epsilon_{27}$ & $\zeta_{27}$ & $\psi_{27}$\\ \hline
$\Sigma^-\pi^-$ & $\sqrt{2}$ & $0$ & $0$ & $0$ & 0 & $4$ & $4$ \\ \hline
\end{tabular}
\end{center}

\begin{center}
\begin{tabular}{|c|c|c|c|c|c|c|c|} \hline
$\Sigma^+_{27}$ & $\alpha_{27}$ & $\beta_{27}$ & $\gamma_{27}$ & $\delta_{27}$ & $\epsilon_{27}$ & $\zeta_{27}$ & $\psi_{27}$\\ \hline
$P\overline{K^0}$ & $-\sqrt{2/5}$ & $\sqrt{3}$ & 0 & $4/\sqrt{5}$ & $-4/\sqrt{5}$ & $4/5\sqrt{5}$ &  $4/5\sqrt{5}$\\ \hline
$\Xi^0K^+$ & $-\sqrt{2/5}$ & 0 & $\sqrt{3}$ & $-4/\sqrt{5}$ & $4/\sqrt{5}$ & $4/5\sqrt{5}$ &  $4/5\sqrt{5}$\\ \hline
$\Lambda \pi^+$ & $\sqrt{3/5}$ & $\sqrt{1/2}$ & $\sqrt{1/2}$ & $-2\sqrt{6/5}$ & $-2\sqrt{6/5}$ & $-2\sqrt{6}/5\sqrt{5}$ & $-2\sqrt{6}/5\sqrt{5}$ \\ \hline
$\Sigma^+ \eta$ & $\sqrt{3/5}$ & $\sqrt{1/2}$ &$\sqrt{1/2}$ & $2\sqrt{6/5}$ & $2\sqrt{6/5}$ & $-2\sqrt{6}/5\sqrt{5}$ & $-2\sqrt{6}/5\sqrt{5}$ \\ \hline
$\Sigma^0 \pi^+$ & 0 & $\sqrt{3/2}$ & $-\sqrt{3/2}$ & $-2\sqrt{2/5}$ & $2\sqrt{2/5}$ & $0$ & $0$ \\ \hline
$\Sigma^+ \pi^0$ & 0 & $-\sqrt{3/2}$ & $\sqrt{3/2}$ & $2\sqrt{2/5}$ & $-2\sqrt{2/5}$ & $0$ & $0$ \\ \hline
\end{tabular}
\end{center}

\begin{center}
\begin{tabular}{|c|c|c|c|c|c|c|c|} \hline
$\Sigma^0_{27}$ & $\alpha_{27}$ & $\beta_{27}$ & $\gamma_{27}$ & $\delta_{27}$ & $\epsilon_{27}$ & $\zeta_{27}$ & $\psi_{27}$\\ \hline
$\Lambda\pi^0$ & $-\sqrt{3/5}$ & $-\sqrt{1/2}$ & $-\sqrt{1/2}$ & $2\sqrt{6/5}$ & $2\sqrt{6/5}$ & $2\sqrt{6}/5\sqrt{5}$ &  $2\sqrt{6}/5\sqrt{5}$\\ \hline
$\Sigma^0\eta$ & $-\sqrt{3/5}$ & $-\sqrt{1/2}$ & $-\sqrt{1/2}$ & $-2\sqrt{6/5}$ & $-2\sqrt{6/5}$ & $2\sqrt{6}/5\sqrt{5}$ &  $2\sqrt{6}/5\sqrt{5}$\\ \hline
$N\overline{K^0}$ & $-\sqrt{1/5}$ & $\sqrt{3/2}$ & 0 & $2\sqrt{2/5}$ & $-2\sqrt{2/5}$ & $2\sqrt{2}/5\sqrt{5}$ &  $2\sqrt{2}/5\sqrt{5}$\\ \hline
$PK^-$ & $\sqrt{1/5}$ & $-\sqrt{3/2}$ & 0 & $-2\sqrt{2/5}$ & $2\sqrt{2/5}$ & $-2\sqrt{2}/5\sqrt{5}$ &  $-2\sqrt{2}/5\sqrt{5}$\\ \hline
$\Xi^-K^+$ & $\sqrt{1/5}$ & 0 & $-\sqrt{3/2}$ & $2\sqrt{2/5}$ & $-2\sqrt{2/5}$ & $-2\sqrt{2}/5\sqrt{5}$ &  $-2\sqrt{2}/5\sqrt{5}$\\ \hline
$\Xi^0K^0$ & $-\sqrt{1/5}$ & 0 & $\sqrt{3/2}$ & $-2\sqrt{2/5}$ & $2\sqrt{2/5}$ & $2\sqrt{2}/5\sqrt{5}$ &  $2\sqrt{2}/5\sqrt{5}$\\ \hline
$\Sigma^-\pi^+$ & $0$ & $\sqrt{3/2}$ & $-\sqrt{3/2}$ & $-2\sqrt{2/5}$ & $2\sqrt{2/5}$ & 0 & 0 \\ \hline
$\Sigma^+\pi^-$ & $0$ & $-\sqrt{3/2}$ & $\sqrt{3/2}$ & $2\sqrt{2/5}$ & $-2\sqrt{2/5}$ & 0 & 0 \\ \hline
\end{tabular}
\end{center}

\begin{center}
\begin{tabular}{|c|c|c|c|c|c|c|c|} \hline
$\Sigma^-_{27}$ & $\alpha_{27}$ & $\beta_{27}$ & $\gamma_{27}$ & $\delta_{27}$ & $\epsilon_{27}$ & $\zeta_{27}$ & $\psi_{27}$\\ \hline
$NK^-$ & $-\sqrt{2/5}$ & $\sqrt{3}$ & 0 & $4/\sqrt{5}$ & $-4/\sqrt{5}$ & $4/5\sqrt{5}$ &  $4/5\sqrt{5}$\\ \hline
$\Xi^-K^0$ & $-\sqrt{2/5}$ & 0 & $\sqrt{3}$ & $-4/\sqrt{5}$ & $4/\sqrt{5}$ & $4/5\sqrt{5}$ &  $4/5\sqrt{5}$\\ \hline
$\Lambda \pi^-$ & $\sqrt{3/5}$ & $\sqrt{1/2}$ & $\sqrt{1/2}$ & $-2\sqrt{6/5}$ & $-2\sqrt{6/5}$ & $-2\sqrt{6}/5\sqrt{5}$ & $-2\sqrt{6}/5\sqrt{5}$ \\ \hline
$\Sigma^- \eta$ & $\sqrt{3/5}$ & $\sqrt{1/2}$ &$\sqrt{1/2}$ & $2\sqrt{6/5}$ & $2\sqrt{6/5}$ & $-2\sqrt{6}/5\sqrt{5}$ & $-2\sqrt{6}/5\sqrt{5}$ \\ \hline
$\Sigma^0 \pi^-$ & 0 & $-\sqrt{3/2}$ & $\sqrt{3/2}$ & $2\sqrt{2/5}$ & $-2\sqrt{2/5}$ & $0$ & $0$ \\ \hline
$\Sigma^- \pi^0$ & 0 & $\sqrt{3/2}$ & $-\sqrt{3/2}$ & $-2\sqrt{2/5}$ & $2\sqrt{2/5}$ & $0$ & $0$ \\ \hline
\end{tabular}
\end{center}

\begin{center}
\begin{tabular}{|c|c|c|c|c|c|c|c|} \hline
$\Lambda_{27}$ & $\alpha_{27}$ & $\beta_{27}$ & $\gamma_{27}$ & $\delta_{27}$ & $\epsilon_{27}$ & $\zeta_{27}$ & $\psi_{27}$\\ \hline
$N\overline{K^0}$ & $\sqrt{3}/2\sqrt{5}$ & $-3/2\sqrt{2}$ & $3/\sqrt{2}$ & $0$ & $0$ & $-4\sqrt{6}/5\sqrt{5}$ &  $-4\sqrt{6}/5\sqrt{5}$\\ \hline
$PK^-$ & $\sqrt{3}/2\sqrt{5}$ & $-3/2\sqrt{2}$ & $3/\sqrt{2}$ & $0$ & $0$ & $-4\sqrt{6}/5\sqrt{5}$ &  $-4\sqrt{6}/5\sqrt{5}$\\ \hline
$\Lambda\eta$ & $-3\sqrt{3}/2\sqrt{5}$ & $3/2\sqrt{2}$ & $3/2\sqrt{2}$ & $0$ & $0$ & $12\sqrt{6}/5\sqrt{5}$ &  $12\sqrt{6}/5\sqrt{5}$\\ \hline
$\Sigma^-\pi^+$ & $-1/2\sqrt{15}$ & $-3/2\sqrt{2}$ & $-3/2\sqrt{2}$ & $0$ & $0$ & $4\sqrt{2}/5\sqrt{15}$ &  $4\sqrt{2}/5\sqrt{15}$\\ \hline
$\Sigma^0\pi^0$ & $-1/2\sqrt{15}$ & $-3/2\sqrt{2}$ & $-3/2\sqrt{2}$ & $0$ & $0$ & $4\sqrt{2}/5\sqrt{15}$ &  $4\sqrt{2}/5\sqrt{15}$\\ \hline
$\Sigma^+\pi^-$ & $-1/2\sqrt{15}$ & $-3/2\sqrt{2}$ & $-3/2\sqrt{2}$ & $0$ & $0$ & $4\sqrt{2}/5\sqrt{15}$ &  $4\sqrt{2}/5\sqrt{15}$\\ \hline
$\Xi^-K^+$ & $\sqrt{3}/2\sqrt{5}$ & $3/\sqrt{2}$ & $-3/2\sqrt{2}$ & $0$ & $0$ & $-4\sqrt{6}/5\sqrt{5}$ &  $-4\sqrt{6}/5\sqrt{5}$\\ \hline
$\Xi^0K^0$ &  $\sqrt{3}/2\sqrt{5}$ & $3/\sqrt{2}$ & $-3/2\sqrt{2}$ & $0$ & $0$ & $-4\sqrt{6}/5\sqrt{5}$ &  $-4\sqrt{6}/5\sqrt{5}$\\ \hline

\end{tabular}
\end{center}

\begin{center}
\begin{tabular}{|c|c|c|c|c|c|c|c|} \hline
$\Xi^{'+}_{27}$ & $\alpha_{27}$ & $\beta_{27}$ & $\gamma_{27}$ & $\delta_{27}$ & $\epsilon_{27}$ & $\zeta_{27}$ & $\psi_{27}$\\ \hline
$\Xi^0\pi^+$ & $1$ & $0$ & $0$ & 0 & $-3\sqrt{2}$ & $2\sqrt{2}$ & $-\sqrt{2}$ \\ \hline
$\Sigma^+\overline{K^0}$ & $1$ & $0$ & $0$  & 0 & $3\sqrt{2}$ &  $2\sqrt{2}$ & $-\sqrt{2}$ \\ \hline
\end{tabular}
\end{center}

\begin{center}
\begin{tabular}{|c|c|c|c|c|c|c|c|} \hline
$\Xi^{'0}_{27}$ & $\alpha_{27}$ & $\beta_{27}$ & $\gamma_{27}$ & $\delta_{27}$ & $\epsilon_{27}$ & $\zeta_{27}$ & $\psi_{27}$\\ \hline
$\Sigma^+K^-$ & $\sqrt{1/3}$ & $0$ & $0$ & 0 & $\sqrt{6}$ &  $2\sqrt{2/3}$ & $-\sqrt{2/3}$ \\ \hline
$\Sigma^0\overline{K^0}$ & $\sqrt{2/3}$ & $0$ & $0$ & 0 & $2\sqrt{3}$ & $4/\sqrt{3}$ & $-2/\sqrt{3}$ \\ \hline
$\Xi^0\pi^0$ & $\sqrt{2/3}$ & $0$ & $0$ & 0 & $-2\sqrt{3}$ & $4/\sqrt{3}$ & $-2/\sqrt{3}$ \\ \hline
$\Xi^-\pi^+$ & $\sqrt{1/3}$ & $0$ & $0$ & 0 & $-\sqrt{6}$ &  $2\sqrt{2/3}$ & $-\sqrt{2/3}$ \\ \hline
\end{tabular}
\end{center}

\begin{center}
\begin{tabular}{|c|c|c|c|c|c|c|c|} \hline
$\Xi^{'-}_{27}$ & $\alpha_{27}$ & $\beta_{27}$ & $\gamma_{27}$ & $\delta_{27}$ & $\epsilon_{27}$ & $\zeta_{27}$ & $\psi_{27}$\\ \hline
$\Sigma^-\overline{K^0}$ & $-\sqrt{1/3}$ & $0$ & $0$ & $0$ & $-\sqrt{6}$ & $-2\sqrt{2/3}$ & $\sqrt{2/3}$ \\ \hline
$\Sigma^0K^-$ & $\sqrt{2/3}$ & $0$ & $0$ & 0 &$2\sqrt{3}$ & $4/\sqrt{3}$ & $-2/\sqrt{3}$ \\ \hline
$\Xi^-\pi^0$ & $\sqrt{2/3}$ & $0$ & $0$ & 0 &$-2\sqrt{3}$ & $4/\sqrt{3}$ & $-2/\sqrt{3}$ \\ \hline
$\Xi^0\pi^-$ & $-\sqrt{1/3}$ & $0$ & $0$ & $0$ & $\sqrt{6}$ & $-2\sqrt{2/3}$ & $\sqrt{2/3}$ \\ \hline
\end{tabular}
\end{center}

\begin{center}
\begin{tabular}{|c|c|c|c|c|c|c|c|} \hline
$\Xi^{'--}_{27}$ & $\alpha_{27}$ & $\beta_{27}$ & $\gamma_{27}$ & $\delta_{27}$ & $\epsilon_{27}$ & $\zeta_{27}$ & $\psi_{27}$\\ \hline
$\Xi^-\pi^-$ & $1$ & $0$ & $0$ & 0 & $-3\sqrt{2}$ & $2\sqrt{2}$ & $-\sqrt{2}$ \\ \hline
$\Sigma^-K^-$ & $1$ & $0$ & $0$  & 0 & $3\sqrt{2}$ &  $2\sqrt{2}$ & $-\sqrt{2}$ \\ \hline
\end{tabular}
\end{center}

\begin{center}
\begin{tabular}{|c|c|c|c|c|c|c|c|} \hline
$\Xi^{0}_{27}$ & $\alpha_{27}$ & $\beta_{27}$ & $\gamma_{27}$ & $\delta_{27}$ & $\epsilon_{27}$ & $\zeta_{27}$ & $\psi_{27}$\\ \hline
$\Xi^0\pi^0$ & $-\sqrt{1/30}$ &  $-3/2$ & $0$ & $\sqrt{3/5}$ &0 & $-1/5\sqrt{15}$ &$14/5\sqrt{15}$ \\ \hline
$\Xi^-\pi^+$ & $\sqrt{1/15}$ &$3/\sqrt{2}$ &  0& $-\sqrt{6/5}$  &0 & $\sqrt{2}/5\sqrt{15}$ & $-14\sqrt{2}/5\sqrt{15}$\\ \hline
$\Xi^0\eta$ & $3/\sqrt{10}$ & $\sqrt{3}/2$ & $-\sqrt{3}$ & $3/\sqrt{5}$ & 0 &  $3/5\sqrt{5}$ & $-42/5\sqrt{5}$ \\ \hline
$\Lambda\overline{K^0}$ & $3/\sqrt{10}$ & $-\sqrt{3}$ & $\sqrt{3}/2$ & $-3/\sqrt{5}$ & 0 & $3/5\sqrt{5}$ & $-42/5\sqrt{5}$\\ \hline
$\Sigma^0\overline{K^0}$ & $-\sqrt{1/30}$ & 0 & $-3/2$ &  $-\sqrt{3/5}$ & 0  & $-1/5\sqrt{15}$ & $14/5\sqrt{15}$ \\ \hline
$\Sigma^+K^-$ & $\sqrt{1/15}$ & 0 & $3/\sqrt{2}$ & $\sqrt{6/5}$ & 0 & $\sqrt{2}/5\sqrt{15}$ & $-14\sqrt{2}/5\sqrt{15}$\\ \hline
\end{tabular}
\end{center}

\begin{center}
\begin{tabular}{|c|c|c|c|c|c|c|c|} \hline
$\Xi^{-}_{27}$ & $\alpha_{27}$ & $\beta_{27}$ & $\gamma_{27}$ & $\delta_{27}$ & $\epsilon_{27}$ & $\zeta_{27}$ & $\psi_{27}$\\ \hline
$\Xi^-\pi^0$ & $\sqrt{1/30}$ &  $3/2$ & 0 &  $-\sqrt{3/5}$ & 0  & $1/5\sqrt{15}$ & $-14/5\sqrt{15}$ \\ \hline
$\Xi^0\pi^-$ & $\sqrt{1/15}$ &$3/\sqrt{2}$ &  0& $-\sqrt{6/5}$  &0 & $\sqrt{2}/5\sqrt{15}$ & $-14\sqrt{2}/5\sqrt{15}$\\ \hline
$\Xi^-\eta$ & $3/\sqrt{10}$ & $\sqrt{3}/2$ & $-\sqrt{3}$ & $3/\sqrt{5}$ & 0 &  $3/5\sqrt{5}$ & $-42/5\sqrt{5}$ \\ \hline
$\Lambda K^-$ & $3/\sqrt{10}$ & $-\sqrt{3}$ & $\sqrt{3}/2$ & $-3/\sqrt{5}$ & 0 &  $3/5\sqrt{5}$ & $-42/5\sqrt{5}$ \\ \hline
$\Sigma^0K^-$ & $\sqrt{1/30}$ & 0 & $3/2$ &  $\sqrt{3/5}$ & 0  & $1/5\sqrt{15}$ & $-14/5\sqrt{15}$ \\ \hline
$\Sigma^-\overline{K^0}$ & $\sqrt{1/15}$ &0 &$3/\sqrt{2}$ &  $\sqrt{6/5}$  &0 & $\sqrt{2}/5\sqrt{15}$ & $-14\sqrt{2}/5\sqrt{15}$\\ \hline
\end{tabular}
\end{center}

\begin{center}
\begin{tabular}{|c|c|c|c|c|c|c|c|} \hline
$\Omega^{0}_{27}$ & $\alpha_{27}$ & $\beta_{27}$ & $\gamma_{27}$ & $\delta_{27}$ & $\epsilon_{27}$ & $\zeta_{27}$ & $\psi_{27}$\\ \hline
$\Xi^0\overline{K^0}$ & $\sqrt{2}$ & $0$ & $0$ & $0$ & 0 & $4$ & $-8$ \\ \hline
\end{tabular}
\end{center}

\begin{center}
\begin{tabular}{|c|c|c|c|c|c|c|c|} \hline
$\Omega^{-}_{27}$ & $\alpha_{27}$ & $\beta_{27}$ & $\gamma_{27}$ & $\delta_{27}$ & $\epsilon_{27}$ & $\zeta_{27}$ & $\psi_{27}$\\ \hline
$\Xi^0K^-$ & $1$ & $0$ & $0$ & $0$ & 0 & $2\sqrt{2}$ & $-4\sqrt{2}$ \\ \hline
$\Xi^-\overline{K^0}$ & $1$ & $0$ & $0$ & $0$ & 0 & $2\sqrt{2}$ & $-4\sqrt{2}$ \\ \hline
\end{tabular}
\end{center}

\begin{center}
\begin{tabular}{|c|c|c|c|c|c|c|c|} \hline
$\Omega^{--}_{27}$ & $\alpha_{27}$ & $\beta_{27}$ & $\gamma_{27}$ & $\delta_{27}$ & $\epsilon_{27}$ & $\zeta_{27}$ & $\psi_{27}$\\ \hline
$\Xi^-K^-$ & $\sqrt{2}$ & $0$ & $0$ & $0$ & 0 & $4$ & $-8$ \\ \hline
\end{tabular}
\end{center}

\end{section}

\begin{section}{Decay Amplitudes of $35$ Pentaquark Multiplet}
\label{35table}

\begin{center}
\begin{tabular}{|c|c|c|}\hline
$\Theta^-_{35},\Theta^0_{35},\Theta^+_{35},\Theta^{++}_{35},\Theta^{+++}_{35}$ & $\beta_{35}$ &$ \gamma_{35}$ \\ \hline
{\rm Stable} & 0 & 0 \\ \hline
\end{tabular}
\end{center}

\begin{center}
\begin{tabular}{|c|c|c|}\hline
$\Delta^{'+++}_{35},\Delta^{'++}_{35},\Delta^{'+}_{35},\Delta^{'0}_{35},\Delta^{'-}_{35},\Delta^{'--}_{35}$ & $\beta_{35}$ &$ \gamma_{35}$ \\ \hline
{\rm Stable} & 0 & 0 \\ \hline
\end{tabular}
\end{center}

\begin{center}
\begin{tabular}{|c|c|c|}\hline
$\Delta^{++}_{35}$ & $\beta_{35}$ &$ \gamma_{35}$ \\ \hline
$P\pi^+$ & $-1$ & $1$ \\ \hline
$\Sigma^+K^+$ &  $1$ & $1$ \\ \hline
\end{tabular}
\end{center}

\begin{center}
\begin{tabular}{|c|c|c|}\hline
$\Delta^{+}_{35}$ & $\beta_{35}$ &$ \gamma_{35}$ \\ \hline
$P\pi^0$ & $\sqrt{2/3}$ & $-\sqrt{2/3}$ \\ \hline
$N\pi^+$ & $-\sqrt{1/3}$& $\sqrt{1/3}$ \\ \hline
$\Sigma^0K^+$ &  $-\sqrt{2/3}$ & $-\sqrt{2/3}$ \\ \hline
$\Sigma^+K^0$ & $\sqrt{1/3}$ & $\sqrt{1/3}$ \\ \hline
\end{tabular}
\end{center}

\begin{center}
\begin{tabular}{|c|c|c|}\hline
$\Delta^{0}_{35}$ & $\beta_{35}$ &$ \gamma_{35}$ \\ \hline
$P\pi^-$ & $\sqrt{1/3}$ & $-\sqrt{1/3}$ \\ \hline
$N\pi^0$ & $\sqrt{2/3}$& $-\sqrt{2/3}$ \\ \hline
$\Sigma^-K^+$ &  $-\sqrt{1/3}$ & $-\sqrt{1/3}$ \\ \hline
$\Sigma^0K^0$ & $-\sqrt{2/3}$ & $-\sqrt{2/3}$ \\ \hline
\end{tabular}
\end{center}

\begin{center}
\begin{tabular}{|c|c|c|}\hline
$\Delta^{-}_{35}$ & $\beta_{35}$ &$ \gamma_{35}$ \\ \hline
$N\pi^-$ & $1$ & $-1$ \\ \hline
$\Sigma^-K^0$ &  $-1$ & $-1$ \\ \hline
\end{tabular}
\end{center}

\begin{center}
\begin{tabular}{|c|c|c|}\hline
$\Sigma^{'++}_{35}$ & $\beta_{35}$ &$ \gamma_{35}$ \\ \hline
$\Sigma^+\pi^+$ & $0$ & $-2\sqrt{6/5}$ \\ \hline
\end{tabular}
\end{center}

\begin{center}
\begin{tabular}{|c|c|c|}\hline
$\Sigma^{'+}_{35}$ & $\beta_{35}$ &$ \gamma_{35}$ \\ \hline
$\Sigma^0\pi^+$ & $0$ & $-2\sqrt{3/5}$ \\ \hline
$\Sigma^+\pi^0$ & $0$ & $-2\sqrt{3/5}$ \\ \hline
\end{tabular}
\end{center}

\begin{center}
\begin{tabular}{|c|c|c|}\hline
$\Sigma^{'0}_{35}$ & $\beta_{35}$ &$ \gamma_{35}$ \\ \hline
$\Sigma^0\pi^0$ & $0$ & $4/\sqrt{5}$ \\ \hline
$\Sigma^-\pi^+$ & $0$ & $-2/\sqrt{5}$ \\ \hline
$\Sigma^+\pi^-$ & $0$ & $-2/\sqrt{5}$ \\ \hline
\end{tabular}
\end{center}

\begin{center}
\begin{tabular}{|c|c|c|}\hline
$\Sigma^{'-}_{35}$ & $\beta_{35}$ &$ \gamma_{35}$ \\ \hline
$\Sigma^0\pi^-$ & $0$ & $2\sqrt{3/5}$ \\ \hline
$\Sigma^-\pi^0$ & $0$ & $2\sqrt{3/5}$ \\ \hline
\end{tabular}
\end{center}

\begin{center}
\begin{tabular}{|c|c|c|}\hline
$\Sigma^{'--}_{35}$ & $\beta_{35}$ &$ \gamma_{35}$ \\ \hline
$\Sigma^-\pi^-$ & $0$ & $2\sqrt{6/5}$ \\ \hline
\end{tabular}
\end{center}

\begin{center}
\begin{tabular}{|c|c|c|}\hline
$\Sigma^{+}_{35}$ & $\beta_{35}$ &$ \gamma_{35}$ \\ \hline
$P\overline{K^0}$ & $-2\sqrt{2/15}$ & $2\sqrt{2/15}$ \\ \hline
$\Xi^0K^+$ & $2\sqrt{2/15}$ & $2\sqrt{2/15}$ \\ \hline
$\Lambda\pi^+$ & $2/\sqrt{5}$ & $-2/\sqrt{5}$ \\ \hline
$\Sigma^0\pi^+$ & $2/\sqrt{15}$ & $0$ \\ \hline
$\Sigma^+\pi^0$ & $-2/\sqrt{15}$& $0$ \\ \hline
$\Sigma^+\eta$ & $-2/\sqrt{5}$ & $-2/\sqrt{5}$ \\ \hline
\end{tabular}
\end{center}

\begin{center}
\begin{tabular}{|c|c|c|}\hline
$\Sigma^{0}_{35}$ & $\beta_{35}$ &$ \gamma_{35}$ \\ \hline
$N\overline{K^0}$ & $-2/\sqrt{15}$ & $2/\sqrt{15}$ \\ \hline
$PK^-$ & $2/\sqrt{15}$ & $-2/\sqrt{15}$ \\ \hline
$\Lambda\pi^0$ & $-2/\sqrt{5}$ & $2/\sqrt{5}$ \\ \hline
$\Sigma^-\pi^+$ & $2/\sqrt{15}$ & $0$ \\ \hline
$\Sigma^+\pi^-$ & $-2/\sqrt{15}$ & $0$ \\ \hline
$\Sigma^0\eta$ & $2/\sqrt{5}$ & $2/\sqrt{5}$ \\ \hline
$\Xi^0K^0$ & $2/\sqrt{15}$ & $2/\sqrt{15}$ \\ \hline
$\Xi^-K^+$ & $-2/\sqrt{15}$ &  $-2/\sqrt{15}$\\ \hline
\end{tabular}
\end{center}

\begin{center}
\begin{tabular}{|c|c|c|}\hline
$\Sigma^{-}_{35}$ & $\beta_{35}$ &$ \gamma_{35}$ \\ \hline
$NK^-$ & $2\sqrt{2/15}$ & $-2\sqrt{2/15}$ \\ \hline
$\Xi^-K^0$ & $-2\sqrt{2/15}$ &  $-2/\sqrt{2/15}$\\ \hline
$\Lambda\pi^-$ & $-2/\sqrt{5}$ & $2/\sqrt{5}$ \\ \hline
$\Sigma^-\pi^0$ & $-2/\sqrt{15}$ & $0$ \\ \hline
$\Sigma^0\pi^-$ & $2/\sqrt{15}$ & $0$ \\ \hline
$\Sigma^-\eta$ & $2/\sqrt{5}$ & $2/\sqrt{5}$ \\ \hline
\end{tabular}
\end{center}

\begin{center}
\begin{tabular}{|c|c|c|}\hline
$\Xi^{'+}_{35}$ & $\beta_{35}$ &$ \gamma_{35}$ \\ \hline
$\Xi^0\pi^+$ & $0$ & $-4/\sqrt{5}$ \\ \hline
$\Sigma^+\overline{K^0}$ & $0$ & $-4/\sqrt{5}$ \\ \hline
\end{tabular}
\end{center}

\begin{center}
\begin{tabular}{|c|c|c|}\hline
$\Xi^{'0}_{35}$ & $\beta_{35}$ &$ \gamma_{35}$ \\ \hline
$\Xi^-\pi^+$ & $0$ & $-4/\sqrt{15}$ \\ \hline
$\Xi^0\pi^0$ & $0$ & $-4\sqrt{2/15}$ \\ \hline
$\Sigma^+K^-$ & $0$ & $-4/\sqrt{15}$\\ \hline
$\Sigma^0\overline{K^0}$ & $0$ & $-4\sqrt{2/15}$ \\ \hline
\end{tabular}
\end{center}

\begin{center}
\begin{tabular}{|c|c|c|}\hline
$\Xi^{'-}_{35}$ & $\beta_{35}$ &$ \gamma_{35}$ \\ \hline
$\Xi^0\pi^-$ & $0$ & $-4/\sqrt{15}$ \\ \hline
$\Xi^-\pi^0$ & $0$ & $4\sqrt{2/15}$ \\ \hline
$\Sigma^-\overline{K^0}$ & $0$ & $-4/\sqrt{15}$ \\ \hline
$\Sigma^0K^-$ & $0$ & $4\sqrt{2/15}$\\ \hline
\end{tabular}
\end{center}

\begin{center}
\begin{tabular}{|c|c|c|}\hline
$\Xi^{'--}_{35}$ & $\beta_{35}$ &$ \gamma_{35}$ \\ \hline
$\Xi^-\pi^-$ & $0$ & $-4/\sqrt{5}$ \\ \hline
$\Sigma^-K^-$ & $0$ & $-4/\sqrt{5}$ \\ \hline
\end{tabular}
\end{center}

\begin{center}
\begin{tabular}{|c|c|c|}\hline
$\Xi^{0}_{35}$ & $\beta_{35}$ &$ \gamma_{35}$ \\ \hline
$\Xi^0\pi^0$ & $-\sqrt{3/10}$ & $\sqrt{1/30}$ \\ \hline
$\Xi^-\pi^+$ & $\sqrt{3/5}$ & $-\sqrt{1/15}$ \\ \hline
$\Xi^0\eta$ & $-3/\sqrt{10}$ & $-3/\sqrt{10}$ \\ \hline
$\Lambda\overline{K^0}$ & $3/\sqrt{10}$ & $-3/\sqrt{10}$ \\ \hline
$\Sigma^0\overline{K^0}$ & $\sqrt{3/10}$ & $\sqrt{1/30}$ \\ \hline
$\Sigma^+K^-$ & $-\sqrt{3/5}$ & $-\sqrt{1/15}$\\ \hline
\end{tabular}
\end{center}

\begin{center}
\begin{tabular}{|c|c|c|}\hline
$\Xi^{-}_{35}$ & $\beta_{35}$ &$ \gamma_{35}$ \\ \hline
$\Xi^-\pi^0$ & $-\sqrt{3/10}$ & $\sqrt{1/30}$ \\ \hline
$\Xi^0\pi^-$ & $-\sqrt{3/5}$ & $\sqrt{1/15}$ \\ \hline
$\Xi^-\eta$ & $3/\sqrt{10}$ & $3/\sqrt{10}$ \\ \hline
$\Lambda K^-$ & $-3/\sqrt{10}$ & $3/\sqrt{10}$ \\ \hline
$\Sigma^0K^-$ & $\sqrt{3/10}$ & $\sqrt{1/30}$ \\ \hline
$\Sigma^-\overline{K^0}$ & $\sqrt{3/5}$ & $\sqrt{1/15}$ \\ \hline
\end{tabular}
\end{center}

\begin{center}
\begin{tabular}{|c|c|c|}\hline
$\Omega^{'0}_{35}$ & $\beta_{35}$ &$ \gamma_{35}$ \\ \hline
$\Xi^0\overline{K^0}$ & $0$ & $-2/\sqrt{6/5}$ \\ \hline
\end{tabular}
\end{center}

\begin{center}
\begin{tabular}{|c|c|c|}\hline
$\Omega^{'-}_{35}$ & $\beta_{35}$ &$ \gamma_{35}$ \\ \hline
$\Xi^0K^-$ & $0$ & $-2\sqrt{3/5}$ \\ \hline
$\Xi^-\overline{K^0}$ & $0$ & $-2\sqrt{3/5}$ \\ \hline
\end{tabular}
\end{center}

\begin{center}
\begin{tabular}{|c|c|c|}\hline
$\Omega^{'--}_{35}$ & $\beta_{35}$ &$ \gamma_{35}$ \\ \hline
$\Xi^-K^-$ & $0$ & $2/\sqrt{6/5}$ \\ \hline
\end{tabular}
\end{center}

\begin{center}
\begin{tabular}{|c|c|c|}\hline
$\Omega^{-}_{35}$ & $\beta_{35}$ &$ \gamma_{35}$ \\ \hline
$\Xi^0K^-$ &  $-2\sqrt{2/5}$ & 0 \\ \hline
$\Xi^-\overline{K^0}$ & $2\sqrt{2/5}$ & 0 \\ \hline
\end{tabular}
\end{center}

\begin{center}
\begin{tabular}{|c|c|c|}\hline
$\Phi^{-}_{35}$, $\Phi^{0}_{35}$& $\beta_{35}$ &$ \gamma_{35}$ \\ \hline
{\rm Stable}& $0$ & $0$ \\ \hline
\end{tabular}
\end{center}

\end{section}

\end{document}